\documentclass[twocolumn,amsmath,amssymb,prb,showpacs]{revtex4}
\usepackage[dvips]{graphicx}
\usepackage{bm}% bold math

\usepackage{pifont}
\usepackage{amssymb}
\usepackage{colordvi}
\usepackage{color}
\definecolor{darkgray}{rgb}{0.3,0.3,0.3}
\definecolor{gray}{rgb}{0.5,0.5,0.5}
\definecolor{yellow}{rgb}{.4,.4,0}
\definecolor{orange}{rgb}{1,0.5,0}
\definecolor{darkgreen}{rgb}{0,0.5,0}
\definecolor{darkblue}{rgb}{0,0,0.5}
\definecolor{darkred}{rgb}{0.5,0,0}
\definecolor{purple}{rgb}{0.45,0,0.35}

\begin{document}
\title{
Fluctuation theorem for heat transport probed by a thermal electrode
}
\author{Y. Utsumi}
\address{Department of Physics Engineering, Faculty of Engineering, Mie University, Tsu, Mie, 514-8507, Japan}

\author{O. Entin-Wohlman}
\affiliation{Physics Department, Ben Gurion University, Beer Sheva 84105, Israel}
\affiliation{Raymond and Beverly Sackler School of Physics and Astronomy, Tel Aviv University, Tel Aviv 69978, Israel}

\author{A. Aharony}
\affiliation{Physics Department, Ben Gurion University, Beer Sheva 84105, Israel}
\affiliation{Raymond and Beverly Sackler School of Physics and Astronomy, Tel Aviv University, Tel Aviv 69978, Israel}

\author{T. Kubo}
\affiliation{Graduate School of Pure and Applied Sciences, University of Tsukuba, Tsukuba, Ibaraki 305-8571, Japan}

\author{Y. Tokura}
\affiliation{Graduate School of Pure and Applied Sciences, University of Tsukuba, Tsukuba, Ibaraki 305-8571, Japan}

\begin{abstract}

We analyze the full-counting statistics of the electric heat current flowing in a two-terminal quantum conductor whose temperature is  probed by a third electrode (``probe electrode"). 
In particular we demonstrate that the cumulant-generating function obeys the fluctuation theorem in the presence of a constant magnetic field. 
The analysis is based on the scattering matrix of the three-terminal junction (comprising of the two electronic terminals and the probe electrode),  and a separation of time scales: it is assumed  that the rapid charge transfer across the conductor and the rapid relaxation of the electrons inside the probe electrode give rise to much slower energy fluctuations in the latter. % that electrode. 
This separation allows for a stochastic treatment of the probe dynamics, and the reduction of the three-terminal setup to an effective two-terminal one. Expressions for the  lowest nonlinear transport coefficients, e.g., the linear-response heat-current noise and the second nonlinear thermal conductance,  are obtained and explicitly shown to preserve the symmetry of the fluctuation theorem for the two-terminal conductor. 
The derivation of our expressions which is based on  the transport coefficients of the three-terminal system explicitly satisfying the fluctuation theorem, requires the full calculations of vertex corrections. 

\end{abstract}

\date{\today}

 \pacs{05.30.-d,72.70+m,73.63.Kv}

\maketitle
\newcommand{\mat}[1]{\mbox{\boldmath$#1$}}
\newcommand{\mtau}{\mbox{\boldmath$\tau$}}

\section{Introduction}

\label{intro}

The recent progress in the research of thermometry, refrigeration~\cite{Giazotto} and heating~\cite{Molenkamp} processes in mesoscopic quantum systems enables one to treat systematically heat-related phenomena of electrons. \cite{Moller,Venkatachalam,Matthews} 
Recent efforts are focused on thermoelectric transport in coherent quantum-conductors coupled to local vibrational modes, \cite{EIA} or to temperature and/or potential-probing electrodes, \cite{Sanchez1} in the linear-response regime and also beyond it.  \cite{SL,Lopez2,Svensson} 
An intriguing question raised in these investigations is the symmetry of the various transport coefficients with respect to time-reversal breaking, in particular under the effect of inelastic interactions induced by  probing electrodes. 
This question is further related to the fluctuation theorem (FT) obeyed by the cumulant-generating function. % in the absence of time-reversal symmetry.

\begin{figure}[ht]
\includegraphics[width=.6 \columnwidth]{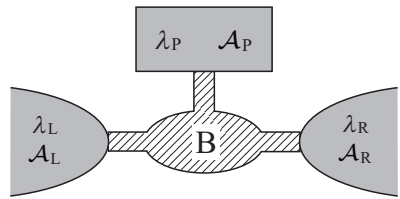}
\caption{
A three-terminal setup. The figure depicts a  quantum conductor    (the elliptical area)  connected to two reservoirs denoted $L$ and $R$, which are held at two different temperatures; 
those are expressed in terms of the affinities ${\cal A}_{L,R} \equiv \beta-\beta_{L,R}$ ($\beta^{-1}$ is the temperature of the entire  system when at equilibrium), see text. 
The  thermal probe (the upper square) is specified by its own affinity ${\cal A}_{P} \equiv \beta-\beta_{P}$, which fluctuates in time. 
Also shown are the three auxiliary ``counting" fields $\lambda_{r}$ ($r=L$, $R$,  and $P$) which measure the energy flowing in and out of electrode $r$. 
The quantum conductor is threaded by a perpendicularly-oriented magnetic field $B$. 
Once the stochastic dynamics of the energy current in and out of the probe electrode is taken care of, the junction becomes an effective two-terminal one.
}
\label{1}
\end{figure}

The prototype setup of coherent thermoelectric transport is a mesoscopic conductor connected to leads kept at various temperatures and a common chemical potential. 
%chemical potentials. 
For clarity, we focus below on a conductor coupled to two electronic reservoirs held at two (different) temperatures, $T_L$ and $T_R$. 
%, but at a common chemical potential. 
Broken time-reversal invariance is induced by a perpendicular magnetic field,  $B$, which affects the orbital motion of the electrons (the much smaller effect of the Zeeman interaction is disregarded). 
Our aim is to investigate the statistics of the heat current flowing in the conductor. 
The conductor is further coupled to a third electrode, designed to measure its temperature (see Fig. \ref{1}). \cite{Engquist,Sivan} 
This measurement is accomplished  by adjusting the temperature $T_{P}$ of this third terminal  so that no net energy is flowing between it and the conductor on the average. 
However, the energy current flowing in or out of the probe electrode fluctuates in time and its distribution depends on details of the coupling between the quantum conductor and the probe (e.g., it is Poissonian for tunnel coupling). 
The fluctuations give rise to stochastic variations in the temperature of the probe. 
These in turn affect the higher cumulants (beyond the first two, i.e., the current and the noise) and the probability distribution of the energy current flowing between the two electronic reservoirs, i.e., the full-counting statistics (FCS). \cite{Levitov,Nazarov}
The problem at hand is therefore to find the cumulant generating-function (CGF) which characterizes this FCS, in the presence of the temperature-measuring probe electrode. 
In other words, to obtain the CGF once the three-terminal setup (where all three electrodes are included on equal footing) is mapped onto an effective two-terminal one in which energy is flowing between the left ($L$) and the right ($R$) reservoirs (see Fig. \ref{1}). 
An ensuing issue is the fate of the FT (imposed on the CGF) under this mapping, in particular when time-reversal invariance is broken. 

A similar situation has been encountered in the statistics of charge currents. 
There, one has to allow for voltage-measuring probes (electrodes whose potential is adjusted as to bar electric currents between them and the conductor) or dephasing probes (which exchange incoherently electrons with the conductor within a narrow energy interval). 
The treatment of the stochastic effects of these electrodes on the CGF is based on time-scales separation. 
For example, the rapid flow of electrons in and out of a voltage-probing electrode results in much slower charge fluctuations there, thus allowing for a stochastic path-integration of the CGF of the full setup (e.g., a three-terminal one) over all configurations of the probe charge, to obtain the reduced CGF of the physical setup (e.g., a two-terminal one). \cite{Pilgram1,Jordan,Pilgram2,Foerster} 
A similar treatment has been carried out for the stochastic temperature and chemical potential fluctuations in an overheated metallic island. \cite{Heikkila,Laakso,Golubev} 

However, to the best of our knowledge there are no studies of the fluctuation theorem (FT)~\cite{EG,Tobiska,FB,FB2,SU,US,Andrieux,Esposito,Campisi,Altland,Lopez,UGMSFS,Kueng,Nakamura,Saira,JW,Agarwalla} in systems coupled to thermal probes.
This paper is devoted to the exploration of this issue. 
In order to map the three-terminal junction of Fig. \ref{1} onto the effective two-terminal one we adopt the stochastic path-integral formalism, \cite{Pilgram1,Jordan,Pilgram2,Foerster} originally devised for describing electric conduction through a chaotic cavity.  
We analyze the FT pertaining to the resulting  effective two-terminal setup. 
In particular we investigate the symmetry relations of the  nonlinear thermal conductances and the linear-response expressions for the corresponding noise correlations and verify that those  obey the universal relations imposed by the FT. \cite{SU} 
Explicit results for the aforementioned energy-transport coefficients are presented by using a triple-quantum-dot junction as an example. 
It allows us to demonstrate the magnetic-field asymmetry induced in the heat transport by the thermal probe (
see e.g. Ref.~\onlinecite{Bedkihal} for related issues), and to confirm that the universal relations imposed by the FT are satisfied.

The FT is a consequence of micro-reversibility and can be considered as a microscopic extension of the second law of thermodynamics.  It can be  expressed in terms of the probability distribution $P_{\tau}(\Delta S)$ for an entropy change $\Delta S$ during a measurement time $\tau$. When time-reversal invariance is broken, say by a  magnetic field $B$,  that probability distribution depends on the latter as well, and the FT  reads 
%----------------------------------------------------------
\begin{align}
\lim_{\tau \to \infty}
\frac{1}{\tau}
\ln 
\frac{P_\tau(\Delta S;B)}{P_\tau(-\Delta S;-B)}
=
I_S
\, ,
\label{ftg}
\end{align}
%----------------------------------------------------------
where  $I_S=\lim_{\tau \to \infty} \Delta S/\tau$ is the entropy flow. 
In a two-terminal junction coupled to two electronic reservoirs held at the same chemical potential  but at different temperatures, $T_{L}$ and $T_{R}$, this flow is \cite{Callen}
%----------------------------------------------------------
\begin{align}
I_{S}^{}
=
I_{E}^{}  (\beta^{}_R - \beta^{}_L)\, ,\label{IS}
\end{align}
%----------------------------------------------------------
where $I_E$ is the energy current, and where $\beta $ denotes the inverse temperature. (We use units in which $e=\hbar=k_{\rm B}=1$ and measure energies from the common chemical potential $\mu=0$,  thus ensuring that the electronic heat current is equivalent to the energy current.) \cite{Callen}
The direction of the currents flow here is out of each electrode. \cite{Buettiker}

Despite its modest form,   the FT Eq.~(\ref{ftg})  is a very powerful relation. It reproduces the linear-response results,  the fluctuation-dissipation theorem and Onsager's reciprocal relations. \cite{Tobiska,FB,FB2,SU,Andrieux,Esposito,Campisi} 
Furthermore, it predicts universal relations among the nonlinear transport coefficients. \cite{FB,FB2,SU,Andrieux} 
A recent experiment has aimed to verify some of these relations, \cite{Nakamura} by comparing the nonlinear conductance of an Aharonov-Bohm interferometer with the noise in the linear-response regime.

The paper is organized as follows. 
We begin in Sec. \ref{FCS} by setting the formal basis of the paper: first we summarize in Sec. \ref{CGFFTa} the probability distribution of the energy currents carried by noninteracting electrons across  a quantum conductor connected to three terminals (i.e. left, right,  and probe electrodes) and its CGF,  together with the symmetries implied by the FT. 
Then, in Sec. \ref{secLangevin}, following the same route taken in Refs. \onlinecite{Pilgram1,Jordan,Pilgram2,Foerster,Heikkila,Laakso,Golubev}, 
we   path-integrate over the stochastic  energy fluctuations in the probe and thus  reduce the three-terminal setup onto an effective two-terminal one. 
We continue in Sec. \ref{FTtwo} by proving that this reduction is consistent with the FT as applied to the reduced two-terminal setup; this is the first main result of this paper. 
In Sec. \ref{steady} we consider the scaled two-terminal CGF at steady state. We continue in Sec. \ref{subtc} by introducing the general scheme for obtaining the transport coefficients when time-reversal symmetry is broken. Then in Sec. 
\ref{subvc} we explain how the transport coefficients of the effective two-terminal junction are obtained from the CGF of the three-terminal one, and introduce the required vertex corrections. 
This analysis allows  us to obtain in Sec. \ref{vercor} the  lowest  nonlinear transport coefficients; this is  the second main result of this paper. 
Finally in Sec. \ref{tp} we apply our theory to the three-terminal triple-quantum dot system. Our results are summarized in Sec.~\ref{summary}.

\section{Full-counting statistics}
\label{FCS}

%----------------------------------------------------------
%----------------------------------------------------------

\subsection{Cumulant-generating function for a quantum conductor coupled to  three terminals  }
\label{CGFFTa}

Figure \ref{1} displays schematically our system:
a
quantum conductor whose temperature ($T_{P}$) is determined by a thermal probe, is subject to a magnetic field $B$ and is attached to  two electronic reservoirs (of temperatures $T_{L}$ and $T_{R}$).
This three-terminal setup is specified by a 3$\times$3 energy  and magnetic-field dependent scattering matrix,  $\mathbf{S}(\omega;B)$, whose elements are the various scattering amplitudes. Those obey  micro-reversibility, 
%----------------------------------------------------------
\begin{align}
S_{rr'}(\omega ;B)=S_{r'r}(\omega; -B) \ , 
\label{micrev}
\end{align}
%----------------------------------------------------------
with $r, r'=L,$ $R$, or $P$. Each of the three terminals is specified by a Fermi distribution at its own temperature, 
%----------------------------------------------------------
\begin{align}
f^{}_{r}(\omega )=[e^{\beta^{}_{r} \omega} +1]^{-1} =[e^{(\beta -{\cal A}^{}_{r})\omega}+1]^{-1}
\ .\label{FF}
\end{align}
%----------------------------------------------------------
%(The energy reference is the common chemical potential of the device, chosen to be zero.)
In the second equality of Eq. (\ref{FF}) we have introduced the affinity ${\cal A}_{r}$ corresponding to the $r$-th reservoir, 
%---------------------------------------------------------- 
\begin{align}
{\mathcal A}^{}_{r}=\beta^{}_{}-\beta^{}_{r} \ ,  
\label{af}
\end{align}
%---------------------------------------------------------- 
where $\beta $ denotes the  common inverse  temperature of the entire junction. These affinities, sometimes called ``thermodynamic forces",  drive the energy currents in the junction.

The statistical properties of the energy transfer are characterized by the probability distribution of the three energy currents, $I_{Er}$, 
emerging from each electrode. 
Alternatively, one may exploit the probability distribution,  $P_{G \tau}( \{ \varepsilon_r \})$,   of the energies accumulated on the three electrodes during the measurement time $\tau$, 
%---------------------------------------------------------
\begin{align}
\varepsilon^{}_{r}\equiv\int_{0}^{\tau}dt I^{}_{Er}(t) \ .\label{er}
\end{align}
%----------------------------------------------------------  
The probability distribution $P_{G \tau}( \{ \varepsilon_r \})$ is also a function of $B$. (For brevity, the explicit dependencies of some of the functions below are suppressed in part of the equations, and is presented when needed for clarity.) The (scaled) cumulant generating-function, ${\cal F}_{G}$, defined in the limit of long measurement times is 
%---------------------------------------------------------- 
\begin{align}
{\cal F}^{}_{G}(\{ \lambda_{r}^{} \})
=
\lim_{\tau\rightarrow\infty}
\frac{1}{\tau}{\rm ln}
{\cal Z}^{}_{G \tau}(\{ \lambda_{r}^{} \})\ ,\label{FG}
\end{align}
%---------------------------------------------------------- 
where ${\cal Z}_{G \tau}$ is the Fourier transform of the probability 
$P_{G \tau}(\{ \varepsilon_{r}^{} \})$ for a finite measurement time $\tau$; 
%----------------------------------------------------------  
\begin{align}
{\cal Z}^{}_{G \tau}(\{ \lambda_{r}^{} \})
=
\int_{-\infty}^{\infty}
\!\!\!
d\varepsilon^{}_{L}
d\varepsilon^{}_{R}
d\varepsilon^{}_{P}
e^{
 i
\sum_{r}  %=L,R,P}
\varepsilon^{}_{r}\lambda^{}_{r}
}
P^{}_{G \tau}( \{ \varepsilon^{}_{r} \})
\ .
\label{defcf}
\end{align}
%----------------------------------------------------------  
We may now examine the symmetries imposed by the FT Eq. (\ref{ftg}) on the CGF ${\cal F}_{G}$. 
The entropy production $\dot{S}$, which in our case is equivalent to the energy current, is given by
 %----------------------------------------------------------
 \begin{align}
\dot{S}=\sum_{r=L,R,P} I^{}_{Er} (\beta-\beta^{}_{r})
=\sum_{r=L,R,P} I^{}_{Er}{\cal A}^{}_{r}\ ,\label{entro}
\end{align}
 %----------------------------------------------------------
where in the last step we have used Eq. (\ref{af}). 
It therefore follows that $\Delta S=\sum_{r}\varepsilon_{r}{\cal A}_{r}$
and consequently, in the limit of long measurement times $\tau\rightarrow\infty$, the FT takes the familiar form
 %----------------------------------------------------------
\begin{align}
P_{G \tau}
( \{ \varepsilon_r \};B)
=
P_{G \tau}
( \{ -\varepsilon_r \};-B)
e^{\sum_{r}  %=L,R,P} 
\varepsilon_r {\mathcal A}_r}
\, .
\label{ftpdf}
\end{align}
%---------------------------------------------------------- 
This, in turn, implies that the CGF obeys  \cite{SU,FB,FB2}
%----------------------------------------------------------
\begin{align}
&{\mathcal F}_G^{}
( \{ \lambda^{}_{r} \}, \{ {\mathcal A}^{}_{r} \};B)
=
{\mathcal F}_G^{}
( \{ -\lambda^{}_{r} +i {\mathcal A}^{}_{r} \}, \{ {\mathcal A}^{}_{r} \}
;-B)
\, .
\label{ftcgfo}
\end{align}
%---------------------------------------------------------- 
Here we stress that the probability distribution $P_{G\tau}$ is for the three-terminal system including the probe terminal. 
As such, it is still an intermediate expression. 
Later in Eq.~(\ref{funcp}), we present the two-terminal probability distribution $P_\tau$ after removing the probe terminal. 
$P_{G \tau}$ and $P_\tau$ are different and should not be confused.

A convenient way to express and calculate the CGF of the energy current~\cite{Kindermann} for {\em noninteracting electrons}  is in terms of the scattering matrix, ${\mathbf S}(\omega;B)$,  
%----------------------------------------------------------
\begin{align} 
{\cal F}_{G}^{}(\{ \lambda_r \})
&=
\int \frac{d\omega}{2\pi} 
\ln{\rm det}
\Bigl ( {\mathbf 1}-{\mathbf f}(\omega ) {\mathbf K} (\mat{\lambda},\omega; B)\Bigr ) \ ,
\label{cgfsm}
\end{align}
%---------------------------------------------------------- 
with the matrix ${\mathbf K}$ given by 
%---------------------------------------------------------- 
\begin{align}
{\mathbf K}(\mat{\lambda},\omega ;B)
&=
{\mat 1}-
e^{i {\mat \lambda} \omega}
{\mathbf S}^{\dagger}_{}(\omega ;B)
e^{-i {\mat \lambda} \omega}
{\mathbf S}(\omega;B) \ .
\label{mk}
\end{align}
%---------------------------------------------------------- 
In Eqs. (\ref{cgfsm}) and (\ref{mk}),  ${\mathbf f}$ is a diagonal matrix of the Fermi functions, 
 %---------------------------------------------------------- 
\begin{align}
{\mathbf f}(\omega )={\rm diag} 
\left\{
f^{}_{L}(\omega),f^{}_{R}(\omega),f^{}_{P}(\omega )
\right\}\ ,
\end{align}  
%----------------------------------------------------------   
and ${\mat \lambda}$ is a diagonal matrix comprising  of the counting fields  
 %----------------------------------------------------------  
\begin{align}
{\mat \lambda}={\rm diag} 
\{
\lambda^{}_{L}, 
\lambda^{}_{R}, 
\lambda^{}_{P} 
\}\ . 
\end{align}
%---------------------------------------------------------- 
Obviously, the CGF  in its form (\ref{cgfsm}) should obey~\cite{FB2} the FT relation Eq. (\ref{ftcgfo}) (see Appendix \ref{ftsm} for details).

\subsection{Stochastic treatment of the probe energy}
\label{secLangevin}

Here we outline the stochastic approach which allows for the path-integration over the slow dynamics of the energy in the probe electrode, and leads to a functional representation for the CGF of the effective two-terminal junction. \cite{Jordan}

As is mentioned above, the temperature of the probe fluctuates in time, i.e., the probe affinity ${\cal A}_{P}$ is time dependent and consequently so is the (instantaneous) probe energy denoted $E(t)$ which is given by
%----------------------------------------------------------
\begin{align}
E(t)
=
\int d \omega 
\rho^{}_P(\omega)
\frac{\omega}
{e^{(\beta-{\mathcal A}_{P}(t)) \omega}+1}
\, , 
\label{probeenergy}
\end{align}
%----------------------------------------------------------
where $\rho_P$ is the electronic density of states in the probe. 
The energy $E(t)$ fluctuates stochastically since the energy current generated by the electrons in the quantum conductor fluctuates. 
Such energy current fluctuations result in non-Gaussian white noise, whose rigorous stochastic calculus has been investigated recently. \cite{Kanazawa} 
In the present paper we adopt a simpler approximation which captures the relevant physics.~\cite{Jordan,Pilgram1,Pilgram2,Foerster,Heikkila,Laakso,Golubev}
This approach relies on the existence of two distinct time scales. 
The faster one pertains to the traveling time of each electron through the conductor and the subsequent relaxation in any of the electrodes. 
The slower one is related to the fluctuations of the energy inside the probe terminal. 

During a time interval $\Delta t$, the energy emitted stochastically from the $r$-th electrode is  
%----------------------------------------------------------
\begin{align}
\Delta \varepsilon_r
=
\int_t^{t+\Delta t}
dt' \, 
I_{Er}(t')
\, . 
\end{align} 
%----------------------------------------------------------
These energy differences obey a joint probability distribution governed by the scaled CGF [see Eqs. (\ref{FG}) and (\ref{defcf})] 
%----------------------------------------------------------
\begin{align}
\label{Pq}
&P^{}_{G \Delta t}( \{ \Delta \varepsilon_r \},
\{ {\mathcal A}_r \} 
;B)\\
&
=
\int_{-\infty}^\infty
\frac{
d \lambda^{}_L
d \lambda^{}_R
d \lambda^{}_{P}
}{(2 \pi)^3}
{\rm e}^{
\Delta t
{\mathcal F}_G
(
\{ \lambda_r \},
\{ {\mathcal A}_r \}
;B)
-i 
\sum_{r} 
\Delta \varepsilon_r \lambda_r
}
\  . \nonumber
\end{align}
%----------------------------------------------------------
However, the time interval $\Delta t$ is chosen in a specific manner designed to single out $\Delta\varepsilon_{P}$ and to make it a stochastic variable. 
Indeed, the key approximation in Refs.~\onlinecite{Jordan,Pilgram1,Pilgram2,Foerster,Heikkila,Laakso,Golubev} is related to the duration of $\Delta t$. 
It should be longer than the time needed for the probe electrode to reach local equilibrium, which is obviously much longer than the time scale characterizing the energy fluctuations in the probe. In other words, $\Delta t$ is much longer than the time required for an electron to relax in the probe electrode. 
The latter time scale is determined for example by electron-electron 
collisions (The probe electrode is in the hot-electron regime).

One next discretizes  the  entire measurement time $\tau$ into 
$N=\tau/\Delta t$ intervals, each of duration $\Delta t$. 
When the probe energy at time $t_n=n\Delta t$ is $E_n$, then after an additional time step it changes to 
%----------------------------------------------------------
\begin{align}
E_{n+1}
=
E_n
+
\Delta \varepsilon_P
\, . 
\label{Langevindis}
\end{align} 
%----------------------------------------------------------
Since $\Delta \varepsilon_P$ is a stochastic variable, 
$E_{n+1}$ is not unique. 
By using the probability distribution of $\Delta \varepsilon_P$, 
Eq. (\ref{Pq}) and Eq.~(\ref{Langevindis}), 
we obtain the conditional joint probability to find the probe energy $E_n$ at time $t=t_n$ and $E_{n+1}$ at $t=t_{n+1}=(n+1)\Delta t $,  accompanied by the energy changes $\Delta \varepsilon_{L n}$ and $\Delta \varepsilon_{R n}$ of the left and right reservoirs
%----------------------------------------------------------
\begin{widetext}
\begin{align}
P^{}_{G \Delta t}
(\Delta \varepsilon^{}_{L n}, 
\Delta \varepsilon^{}_{R n}, 
E^{}_{n+1}-E^{}_n,
{\mathcal A}^{}_L, 
{\mathcal A}^{}_R, 
{\mathcal A}^{}_{P}( [E^{}_{n+1}+E^{}_n]/2 )
;B)
\, , 
\label{conpro} 
\end{align}
\end{widetext}
%----------------------------------------------------------
where we have assumed that during $\Delta t$ 
the probe affinity is determined by  Eq.~(\ref{probeenergy}) with the average  energy at times $t_{n+1}$ and $t_n$,   ${\mathcal A}^{}_{P}( [E^{}_{n+1}+E^{}_n]/2 )$. 
Although from the viewpoint of  causality it would be more reasonable to use instead ${\mathcal A}^{}_{P}(E^{}_n)$, 
this mid-point rule is convenient for proving the FT (see Sec.~\ref{FTtwo}). 

It follows from the above that only two adjacent events determine the probe dynamics, making it a Markov process. 
Hence, the probability distribution  for the energies 
$\varepsilon_{L/R}=\sum_{j=0}^{N-1} \Delta \varepsilon_{L/Rj}$
to emerge from the left and right reservoirs 
during the measurement time $\tau$ is 
%----------------------------------------------------------
\begin{widetext}
\begin{align}
P^{}_{\tau}(\varepsilon^{}_L,\varepsilon^{}_R,{\mathcal A}^{}_L,{\mathcal A}^{}_R;B)
&=
\int 
\Big (\prod ^{N-1}_{j=0}d \Delta \varepsilon^{}_{Lj} 
d \Delta \varepsilon^{}_{Rj}
\Big )\Big (\prod_{j=0}^{N}d E^{}_{j}
\Big )\delta ( \varepsilon^{}_L - \sum_{j=0}^{N-1} \Delta \varepsilon^{}_{L j} )
\delta ( \varepsilon^{}_R - \sum_{j=0}^{N-1} \Delta \varepsilon^{}_{R j} )
\nonumber \\
& \times
P^{}_{G \Delta t}(\Delta \varepsilon^{}_{L N-1},\Delta \varepsilon^{}_{R N-1},E^{}_{N}-E^{}_{N-1},{\mathcal A}^{}_L,{\mathcal A}^{}_R,{\mathcal A}^{}_{P}([E^{}_{N}+E^{}_{N-1}]/2);B)
\cdots
\nonumber \\
& \times 
P^{}_{G \Delta t}(\Delta \varepsilon^{}_{L0},\Delta \varepsilon^{}_{R0},E^{}_{1}-E^{}_{0},{\mathcal A}^{}_L,{\mathcal A}^{}_R,{\mathcal A}_{P}^{}([E^{}_{1}+E^{}_{0}]/2);B)
\, 
p^{}_P(E_{0})
\, .
\label{funcp}
\end{align}
%----------------------------------------------------------
Here 
$p_P(E_0)$ 
is the equilibrium distribution probability of the probe energy at an initial time $t_0$. 
The second and third lines of the right-hand side of Eq.~(\ref{funcp}) express the probability to find a path in  energy space, 
$E_0, E_1, \cdots, E_N$, 
at 
$t_0, t_1, \cdots, t_N$. 
The characteristic function is the Fourier transform of Eq. (\ref{funcp}),
%----------------------------------------------------------
\begin{align}
&{\mathcal Z}_\tau
( \{ \lambda^{}_r \}, \{ {\mathcal A}^{}_r \} ;B)
=
\int 
d \varepsilon^{}_L 
d \varepsilon^{}_R 
{e}^{i \lambda^{}_L \varepsilon^{}_L + i \lambda^{}_R \varepsilon^{}_R}
P^{}_\tau(\{ \lambda^{}_r \}, \{ {\mathcal A}^{}_r \};B)
\nonumber\\
&=
\int\Big ( \prod^{N-1}_{j=0}\frac{d \lambda^{}_{{P}j}}{2\pi}\prod^{N}_{j=0} 
d E^{}_{j} \Big )
p^{}_P(E_{0})
{e}^{-i \sum_{j=0}^{N-1} 
\lambda^{}_{{P} j} (E^{}_{j+1}-E^{}_{j})
+\sum_{j=0}^{N-1}
\Delta t {\mathcal F}_G(\lambda^{}_L,\lambda^{}_R,\lambda^{}_{{P} j},
{\mathcal A}^{}_L,{\mathcal A}^{}_R,{\mathcal A}^{}_{P}( [E^{}_{j+1}+E^{}_{j}]/2 )
;B)
}
\,  ,
\label{CFdiscre}
\end{align}
%----------------------------------------------------------
\end{widetext}
%----------------------------------------------------------
where now    the arguments 
$\{ \lambda_r \}$ and $\{ {\mathcal A}_r \}$ refer to
the counting fields and affinities of solely the left and right electronic reservoirs.  

In the continuum   limit, 
$\Delta t \to 0$, 
the characteristic function becomes 
%----------------------------------------------------------
\begin{align}
{\mathcal Z}_\tau(
\{ \lambda^{}_r \},
\{ {\mathcal A}^{}_r \}
;B)
=
\int {\mathcal D} [\lambda^{}_{P},E] 
\, 
{\rm e}^{i {\mathcal S}}
p^{}_P(E(0))
\, ,
\label{CFST}
\end{align}
%----------------------------------------------------------
where $\int {\mathcal D} [\lambda^{}_{P},E]$ 
means the functional integration over $\lambda^{}_{P}(t)$ and $E(t)$.
The Martin-Siggia-Rose action~\cite{Kamenevbook,Heikkila,Laakso} ${\mathcal S}$  is given by 
%----------------------------------------------------------
\begin{align}
i 
{\mathcal S} 
&=
-
\int_0^\tau dt 
\Big [
i \lambda^{}_{P}(t) \dot{E}(t)
\nonumber \\
&
-
{\mathcal F}^{}_G
(\lambda^{}_L,\lambda^{}_R,\lambda^{}_{P}(t)
,{\mathcal A}^{}_L,{\mathcal A}^{}_R,
{\mathcal A}^{}_{P}(E(t));B)
\Big ]
\, . 
\label{action}
\end{align}
%----------------------------------------------------------

\subsection{Fluctuation theorem for the reduced two-terminal system}
\label{FTtwo}

The probability distribution of the effective two-terminal junction, Eq.  (\ref{funcp}),  is based on the assumption that the  dynamics of the energy flow in and  out of  the probe is slow and can be treated stochastically.
Therefore, it is not {\it a priori} obvious that the CGF thus derived obeys the fluctuation theorem. Here we prove that it does.  

The proof begins with Eqs. (\ref{ftpdf}) and (\ref{conpro}) 
which yield the extended form of the local detailed balance,~\cite{GUMS} 
or the detailed fluctuation theorem,~\cite{Jarzynski} 
%----------------------------------------------------------
\begin{widetext}
\begin{align}
&P^{}_{G \Delta t}(\Delta \varepsilon^{}_{L j-1},\Delta \varepsilon^{}_{R j-1},E^{}_{j}-E^{}_{j-1},{\mathcal A}^{}_L,{\mathcal A}^{}_R,{\mathcal A}^{}_{P}([E^{}_{j}+E^{}_{j-1}]/2);B)
\\
&=
P^{}_{G \Delta t}(-\Delta \varepsilon^{}_{L j-1},-\Delta \varepsilon^{}_{R j-1},E^{}_{j-1}-E^{}_{j},{\mathcal A}^{}_L,{\mathcal A}^{}_R,{\mathcal A}^{}_{P}([E^{}_{j-1}+E^{}_{j}]/2);-B)
e^{\sum_{r=L,R} \Delta \varepsilon_{r \, j-1} {\mathcal A}_r} 
p^{}_P(E^{}_{j})/p^{}_P(E^{}_{j-1})
\, . \nonumber
\end{align}
%----------------------------------------------------------
Here we have  imposed the first law of thermodynamics for  the fluctuating energy, 
%\cite{Sekimoto} 
%----------------------------------------------------------
\begin{align}
{\mathcal A}^{}_{P}( (E^{}_{j+1}+E^{}_{j})/2 ) 
(E^{}_{j+1}-E^{}_{j})
=
\ln p^{}_P(E^{}_{j+1})
-
\ln p^{}_P(E^{}_{j})
+
{\mathcal O}(\Delta t^2)
\, , 
\end{align}
%----------------------------------------------------------
and introduced  the instantaneous equilibrium probability,  $p_P(E_j)$, 
to find the probe energy $E_j$ at time $t_j$. 
It then follows  from Eq.~(\ref{funcp}) that
%----------------------------------------------------------
\begin{align}
&P^{}_{\tau}(\{ \varepsilon_r \},\{ {\mathcal A}_r \};B)
=
\int 
\Big (\prod_{j=0}^{N-1}d \Delta \varepsilon^{}_{L j} %\cdots d \Delta \varepsilon^{}_{L0}
d \Delta \varepsilon^{}_{R j}\Big ) % \cdots d \Delta \varepsilon^{}_{R0}
\Big (\prod_{j=0}^{N}d E^{}_j\Big ) % \cdots d E^{}_0
\, 
\delta ( \varepsilon^{}_L - \sum_{j=0}^{N-1} \Delta \varepsilon^{}_{Lj} )
\, 
\delta ( \varepsilon^{}_R - \sum_{j=0}^{N-1} \Delta \varepsilon^{}_{Rj} )
\nonumber \\ & \times 
e^{\sum_{j=0}^{N-1} 
(
\Delta \varepsilon^{}_{L j} {\mathcal A}^{}_L
+
\Delta \varepsilon^{}_{R j} {\mathcal A}^{}_R
)
}
\, 
P^{}_{\Delta t}(-\Delta \varepsilon^{}_{L0},-\Delta \varepsilon^{}_{R0},E^{}_0-E^{}_1,
{\mathcal A}^{}_L,{\mathcal A}^{}_R,
{\mathcal A}^{}_P([E^{}_0+E^{}_1]/2);-B)
\cdots
\nonumber \\ & \times 
P^{}_{\Delta t}(-\Delta \varepsilon^{}_{L N-1},-\Delta \varepsilon^{}_{R N-1},E^{}_{N-1}-E^{}_{N},{\mathcal A}^{}_L,{\mathcal A}^{}_R,{\mathcal A}^{}_{P}([E^{}_{N-1}+E^{}_{N}]/2);-B)
\, 
p^{}_P(E^{}_{N})
\nonumber\\
&=
P^{}_{\tau}(\{ -\varepsilon^{}_r \},\{ {\mathcal A}^{}_r \};-B)
e^{\varepsilon^{}_L {\mathcal A}^{}_L + \varepsilon^{}_R {\mathcal A}^{}_R}
\, , \ \ \ r=L,R\ .
\label{ftp}
\end{align}
\end{widetext}
%----------------------------------------------------------
Upon Fourier transforming this expression, one finds   that the characteristic function obeys the fluctuation theorem,
%----------------------------------------------------------
\begin{align}
&\mathcal{ Z}^{}_\tau(\{ \lambda^{}_r \}, \{ {\mathcal A}^{}_r \};B)
=
\mathcal{ Z}^{}_\tau(\{ -\lambda^{}_r+i {\mathcal A}^{}_r \},\{ {\mathcal A}^{}_r \};-B)
\, ,
\label{ttftz}
\end{align}
%----------------------------------------------------------
similarly to Eq.~(\ref{ftpdf}). 
Thus we have demonstrated that  the stochastic path-integral treatment  of the temperature probe is consistent with the FT. 
This proof, which is similar to the one given \cite{UGMSK} for the work fluctuation theorem of  an $LC$ circuit coupled to a quantum conductor,  \cite{com} is one of the main results of our paper.

\subsection{The steady state}
\label{steady}

To perform the functional integral of Eq. (\ref{CFST}), we adopt the saddle-point approximation, \cite{Pilgram1} for which
$\delta {\mathcal S}/\delta (i \lambda_P(t))=\delta {\mathcal S}/\delta E(t)=0$, and consequently 
%----------------------------------------------------------
\begin{align}
\dot{E}
=
\frac{
\partial {\mathcal F}^{}_G
}{
\partial (i \lambda_P)
}
\, ,
\;\;\ \ 
i 
\dot{\lambda}_P
=
-
\frac{\partial {\mathcal F}_G}
{\partial E}
\, . 
\label{HE0}
\end{align}
%----------------------------------------------------------
These equations are analogous to Hamilton's equations of motion upon regarding 
$i \lambda_P$ as the `momentum', 
$E$ as the `coordinate',  and 
${\mathcal F}_G$ as the `Hamiltonian' 
(see Sec. 4 in Ref.~\onlinecite{Kamenevbook}). 
When $\lambda_L=\lambda_R=0$, a steady-state solution satisfying $\dot{\lambda}_P=\dot{E}=0$ corresponds to a saddle-point of the `Hamiltonian' residing on the $E$-axis satisfying $i \lambda_P=0$. 
%(see e.g. Sec. 4 in Ref.~\onlinecite{Kamenevbook}).
However, for $\lambda_L, \lambda_R \neq 0$, which we are considering in the present paper, it is not the case. 
It is convenient to use ${\mathcal A}_P$ instead of $E$ by using Eq.~(\ref{probeenergy}). 
Then Eqs.~(\ref{HE0}) can be rewritten as
%----------------------------------------------------------
\begin{align}
T^{2}_P \, 
C^{}_P \, 
\dot{{\mathcal A}_P}
=
\frac{
\partial {\mathcal F}^{}_G
}{
\partial (i \lambda_P)
}
\, ,
\;\;\ \ 
T^{2}_P \, 
C^{}_P \, 
i\dot{\lambda}_P
=
-
\frac{\partial {\mathcal F}_G}
{\partial {\mathcal A}_P}
\, . 
\label{HE}
\end{align}
%----------------------------------------------------------
Here $T_{P}\equiv(\beta - {\mathcal A}_P)^{-1}$ is the (finite) probe temperature and $C_P=\partial E/\partial T_P$ is its heat capacitance. 
Therefore, when the heat capacitance $C_P$ is finite, a steady state can be reached and corresponding solutions ${\lambda}_P^*$ and ${\mathcal A}_{P}^*$ (which are purely imaginary and purely real, respectively) can be calculated from 
%----------------------------------------------------------
\begin{align}
&
\frac{\partial}{\partial \lambda_{P}^*}
{\mathcal F}_G(\lambda_L,\lambda_R,\lambda_{P}^*,{\mathcal A}_L,{\mathcal A}_R,{\mathcal A}_{P}^*;B)
\nonumber \\
&=
\frac{\partial}{\partial {\mathcal A}_{P}^*}
{\mathcal F}_G(\lambda_L,\lambda_R,\lambda_{P}^*,{\mathcal A}_L,{\mathcal A}_R,{\mathcal A}_{P}^*;B)
=
0
\, . 
\label{sadeqsb}
\end{align}
%----------------------------------------------------------
The scaled CGF of the two-terminal junction, 
%----------------------------------------------------------
\begin{align}
{\mathcal F}
(\{ \lambda^{}_r \} ,\{ {\mathcal A}^{}_r \};B)
&=
\lim_{\tau \to \infty}
\frac{\ln {\mathcal Z}^{}_\tau 
(\{ \lambda^{}_r \} ,\{ {\mathcal A}^{}_r \};B)
}{\tau}
\, , 
\label{fscgf}
\end{align}
%----------------------------------------------------------
is then related to the one of the three-terminal junction upon using in the expression for the latter the  saddle-point approximation values
%----------------------------------------------------------
\begin{align}
&{\mathcal F}(\{ \lambda^{}_r \},\{ {\mathcal A}^{}_r \};B)
\nonumber \\
&=
{\mathcal F}^{}_{\rm G}
(\lambda^{}_L,\lambda^{}_R,\lambda_{P}^*,
{\mathcal A}^{}_L,{\mathcal A}^{}_R,{\mathcal A}_{P}^*;B)
\, . 
\label{sadapp}
\end{align}
%----------------------------------------------------------
Equation (\ref{sadapp}) is a large-deviation function, \cite{Touchette} which maximizes the probability to find zero net current through the probe electrode by properly choosing ${\mathcal A}_P$. 
The characteristic function is then approximately given by
${\mathcal Z}_\tau \approx \exp(\tau {\mathcal F})$. 

As a consequence of energy conservation, the two-terminal scaled CGF (\ref{sadapp}) is a function of only the {\em difference} between the  left and the  right counting fields,  
$\lambda = \lambda_L - \lambda_R$. 
To see this, 
we return to the
three-terminal scaled CGF Eq.  (\ref{cgfsm}) and note that, again as a result of energy conservation,  it is invariant under a common shift of the three counting fields, 
%----------------------------------------------------------
\begin{align}
\lambda_{r} \rightarrow \lambda_{r}+\delta\lambda_{}\ . 
\label{cash}
\end{align}
%----------------------------------------------------------
This means that it can be expressed as a function of {\em two} counting fields.  
In addition, $\lambda_P$ is a `floating' variable determined by the saddle-point condition Eq.~(\ref{sadeqsb}), which means that a shift of $\lambda_P$ alone, 
%----------------------------------------------------------
\begin{align}
\lambda_{P} \rightarrow \lambda_{P}+\delta\lambda_{P} \ , 
\label{cashp}
\end{align}
%----------------------------------------------------------
cannot change the scaled CGF belonging to the effective two-terminal junction.

Taking advantage of the invariance under the shifts,   
Eqs. (\ref{cash}) and (\ref{cashp}), 
and rewriting the three-terminal CGF as a function of two independent counting field $\lambda$ and $\lambda_P$ and the corresponding two affinities ${\cal A}$ and ${\cal A}_{P}$, we obtain
%----------------------------------------------------------
\begin{align}
&{\mathcal F}^{}_G(\lambda,\lambda^{}_P,{\mathcal A},{\mathcal A}^{}_P;B,x)
\nonumber \\
&\equiv
{\mathcal F}^{}_G( (1+x) \lambda, x \lambda, \lambda^{}_P, 
(1+x) {\mathcal A}, x {\mathcal A}, {\mathcal A}^{}_P
;B)
\, . 
\label{choi}
\end{align}
%------------------------------------------------------------
Here we have introduced the 
parameter $x$, which measures the asymmetry in the inverse-temperature drop between the left and right electrodes, 
%---------------------------------------------------------- 
\begin{align}
{\mathcal A}^{}_{L}
=
(1+x) \, 
{\mathcal A}^{}_{}
\, , \ \ 
{\mathcal A}^{}_{R}
=
x \, 
{\mathcal A}^{}_{}
\,  \ . 
\label{fixaf}
\end{align}
%----------------------------------------------------------
(This parameter is dictated by the details of the experimental setup.)
Note that the form Eq. (\ref{choi}) for the CGF of the three-terminal setup satisfies the FT,
%----------------------------------------------------------
\begin{align}
&
{\mathcal F}_G
(\lambda,\lambda_P,{\mathcal A},{\mathcal A}_P;B,x)
\nonumber \\
&=
{\mathcal F}_G(-\lambda + i {\mathcal A},
-\lambda_P + i {\mathcal A}_P,
{\mathcal A},{\mathcal A}_P;-B,x)
\label{ftfg}
\, . 
\end{align}
%----------------------------------------------------------
The second advantage of Eq. (\ref{choi}) is that it can be used to simplify the saddle-point condition. By shifting
the counting fields,  $\lambda_r \to \lambda_r -\lambda_R + x \lambda$,  Eq. (\ref{sadeqsb}) becomes
%----------------------------------------------------------
\begin{align}
&
\frac{\partial}{\partial \lambda_{P}^*}
{\mathcal F}_G(\lambda,\lambda_{P}^* -\lambda_R + x \lambda,{\mathcal A},{\mathcal A}_{P}^*;B,x)
\nonumber \\
&=
\frac{\partial}{\partial {\mathcal A}_{P}^*}
{\mathcal F}_G(\lambda,\lambda_{P}^* -\lambda_R + x \lambda,{\mathcal A},{\mathcal A}_{P}^*;B,x)
=
0
\, , 
%\label{sadeqs}
\end{align}
and consequently upon choosing
$\lambda_{P}^* \to \lambda_{P}^* -\lambda_R + x \lambda$, 
we obtain
%----------------------------------------------------------
\begin{align}
&
\frac{\partial}{\partial \lambda_{P}^*}
{\mathcal F}_G(\lambda,\lambda_{P}^*,{\mathcal A},{\mathcal A}_{P}^*;B,x)
\nonumber \\
&=
\frac{\partial}{\partial {\mathcal A}_{P}^*}
{\mathcal F}_G(\lambda,\lambda_{P}^*,{\mathcal A},{\mathcal A}_{P}^*;B,x)
=
0
\, . 
\label{sadeqsd}
\end{align}
%----------------------------------------------------------
The corresponding CGF of the effective two-terminal  junction is then  
%----------------------------------------------------------
\begin{align}
{\mathcal F}(\lambda,{\mathcal A};B,x)
=
{\mathcal F}_{\rm G}
(
\lambda,\lambda_{P}^*,
{\mathcal A},{\mathcal A}_{P}^*;B,x)
\, , 
\label{sadappd}
\end{align}
%----------------------------------------------------------
which depends on a single counting field $\lambda$. 
The two-terminal FT Eq. (\ref{ttftz}) can now be  expressed as
%----------------------------------------------------------
\begin{align}
{\mathcal F}(\lambda,{\mathcal A};B,x)
=
{\mathcal F}(-\lambda + i {\mathcal A},{\mathcal A};-B,x)
\, . 
\label{ttftf}
\end{align}
%----------------------------------------------------------
Equations (\ref{sadeqsd}) and (\ref{sadappd}) are the starting point of the following calculations.

\section{transport coefficients and vertex corrections}

\subsection{Transport coefficients}
\label{subtc}

Once the CGF Eq. (\ref{sadappd}) is found, the various cumulants are obtained upon differentiating it with respect to the counting field and the affinity, 
%----------------------------------------------------------
\begin{align}
\langle \! \langle \delta I_E^{j} \rangle \! \rangle
&=
\left. 
\frac{
\partial^{j}
{\mathcal F}
(\lambda,{\mathcal A};B,x)
}
{\partial (i \lambda)^{j} 
}
\right|_{\lambda=0}
=
\sum_{k=0}^\infty
L^{j}_{k}
\frac{
{\mathcal A}^k
}{k!}
\, ,
\end{align}
%----------------------------------------------------------
where the transport coefficients of the reduced two-terminal junction are given by
%----------------------------------------------------------
\begin{align}
{L}^{j}_{k}
\equiv
\left.
\frac{d^{j+k} {\mathcal F}}
{
d (i \lambda)^j
d {\mathcal A}^k_{} 
}
\right|_{\lambda={\mathcal A}=0}
\, . 
\label{nontracoecha}
\end{align}
%----------------------------------------------------------
For example, $L^{1}_{1}$ corresponds to the linear-response thermal conductivity, while $L^{2}_{0}$ is its noise in equilibrium.

In the presence of a magnetic field, it is convenient to introduce symmetrized/anti-symmetrized forms of the transport coefficients,  
%----------------------------------------------------------
\begin{align}
L^{j}_{k,\pm}
=
L^{j}_{k}(B) \pm L^{j}_{k}(-B) 
\, . \label{sym}
\end{align}
%----------------------------------------------------------
By combining the two-terminal FT (\ref{ttftf}) with the definition  (\ref{nontracoecha}) one obtains certain universal relations among the nonlinear transport coefficients $L^{i}_{j}$, which are valid out of equilibrium. \cite{SU}
Relevant relations are summarized  in Appendix  \ref{nonlineonsager}. 

In %the next section 
Sec.~\ref{subvc} 
we express the  two-terminal transport coefficients using those of the  three-terminal junction.  %transport coefficients. 
The latter are derivatives of the CGF Eq. (\ref{choi}), 
%----------------------------------------------------------
\begin{align}
L^{j \ell}_{k m}(B)
\equiv
\left.
\frac{\partial^{j+k+\ell +m} {\mathcal F}_G
%(\lambda,\lambda_P,{\mathcal A},{\mathcal A}_P;B,x)
}
{
\partial (i \lambda)^j
\partial {\mathcal A}^k
\partial (i \lambda_{P})^\ell
\partial {\mathcal A}_{P}^m
}
\right|_{\lambda=\lambda_{P}={\mathcal A}={\mathcal A}_{P}=0}
\, .
\label{nltrc}
\end{align}
%----------------------------------------------------------
By combining the three-terminal FT (\ref{ftfg}) with the definition (\ref{nltrc}), one obtains certain universal relations among the nonlinear transport coefficients $L^{ik}_{j\ell}$, which are also summarized  in Appendix  \ref{nonlineonsager}.

There is a subtle point related to the choice of the ``coordinates", i.e.,  the counting fields and the affinities (see Appendix    \ref{mcc}).   
As discussed above, the three-terminal transport coefficients are invariant under  shifts of the counting fields,  Eqs. (\ref{cash}) and (\ref{cashp}). 
This redundancy results in different but equivalent expressions for the two-terminal transport coefficients. 
The situation  is similar to what happens in the presence of  gauge fields, where different choices of the gauge result in apparently different expressions, which are, in fact, identical. 
Previous research has exploited the minimal-correlation coordinate~\cite{Jordan} (see Appendix \ref{mcc}), which simplifies 
drastically the calculations at the price of expressions which do not explicitly obey the FT and consequently miss symmetries among the transport coefficients.

\subsection{Vertex corrections}
\label{subvc}

While the procedure outlined above for determining the transport coefficients of the effective two-terminal junction  is seemingly straightforward, 
it is not free of certain pitfalls.
As can be seen from the left-hand side of Eq. (\ref{sadappd}), small variations of the counting field and  affinity of the two-terminal junction, $\lambda$ and ${\mathcal A}$, 
lead to small shifts  in the saddle-point values of the counting field and affinity of the probe, 
$\lambda_P^*$ and ${\mathcal A}_P^*$. These shifts,  in turn,  give rise to corrections in the transport coefficients of the two-terminal junction, i.e. vertex corrections  [see
Eq.~(\ref{vc2}) below].  
Here we outline the derivation of the first few cumulants taking into account these vertex corrections.
Technically, the  procedure we follow is identical to the one performed in the self-consistent $\Phi$-derivable approximation~\cite{Baym,UEUA} and the saddle-point approximation in the Schwinger-Keldysh path-integral approach. \cite{US}

In order to keep the equations compact, we introduce the shorthand notations 
$a_{c}={\mathcal A}$, 
$a_q = i \lambda$, 
and 
$v_c={\mathcal A}_{P}$, 
$v_q = i \lambda_{P}$. 
In terms of these, the saddle-pint equations (\ref{sadeqsd}) become
%----------------------------------------------------------
\begin{align}
\partial {\mathcal F}^{}_G/
\partial v^{}_\alpha=0 \ ,
\label{eq:saddlepoint1}
\end{align}
%----------------------------------------------------------
where $\alpha=c,q$. 
The complete derivative of Eq. 
(\ref{eq:saddlepoint1}) with respect to $a_\gamma$ ($\gamma=c,q$) is
%----------------------------------------------------------
\begin{align}
\frac{d}{d a_\gamma}
\frac{\partial {\mathcal F}^{}_G}{\partial v_\alpha}
=
\frac{\partial^2 {\mathcal F}^{}_G}
{\partial v_\alpha \partial a_\gamma}
+
\sum_{\alpha'=c,q}
\frac{\partial^2 {\mathcal F}^{}_G}
{\partial v^{}_\alpha \partial v^{}_{\alpha'}}
\frac{d v^{}_{\alpha'}}{d a_\gamma}
=0
\, .
\end{align}
%----------------------------------------------------------
It therefore follows that
%----------------------------------------------------------
\begin{align}
\frac{d v^{}_{\alpha}}{d a_\gamma}
=
\sum_{\alpha'=c,q}
U^{}_{\alpha \alpha'}
\frac{\partial^2 {\mathcal F}^{}_G}
{\partial a_\gamma \, \partial v^{}_{\alpha'}}
\, ,
\label{evc}
\end{align}
%----------------------------------------------------------
where the  matrix $U$ obeys
%----------------------------------------------------------
\begin{align}
\sum_{\alpha'=c,q}
U^{}_{\alpha \alpha'}
\frac{\partial^2 {\mathcal F}^{}_G}
{\partial v^{}_{\alpha'} \partial v^{}_{\alpha''}}
=
-
\delta^{}_{\alpha \alpha''}
\, ,
\;\;\;\;
U^{}_{\alpha \alpha''}
=
U^{}_{\alpha'' \alpha}
\, . 
\label{u}
\end{align}
%----------------------------------------------------------
Furthermore, the partial derivatives of $U$ are given by
%----------------------------------------------------------
\begin{align}
\frac{\partial U^{}_{\alpha \alpha'} }{\partial a^{}_{\alpha''}}
&=
\sum_{\gamma,\gamma'=c,q}
U^{}_{\alpha \gamma}
\frac{\partial^3 {\mathcal F}^{}_G}{\partial a^{}_{\alpha''} \, \partial v^{}_{\gamma} \, \partial v^{}_{\gamma'}}
U^{}_{\gamma' \alpha'}
\, ,
%^\label{duda}
\nonumber\\
\frac{\partial U^{}_{\alpha \alpha'} }{\partial v^{}_{\alpha ''}}
&=
\sum_{\gamma,\gamma'=c,q}
U_{\alpha \gamma}
\frac{\partial^3 {\mathcal F}^{}_G}{\partial v^{}_{\alpha'' }\, \partial v^{}_{\gamma} \, \partial v^{}_{\gamma'}}
U^{}_{\gamma' \alpha'}
\, ,
\label{dudv}
\end{align}
%----------------------------------------------------------
which can be verified by differentiating the first of   Eqs. (\ref{u}).

We can now  obtain the first derivative of the CGF of the two-terminal junction in terms of derivatives of the CGF of the three-terminal one, 
%----------------------------------------------------------
\begin{align}
\frac{d {\cal F}}{d a_\gamma}
=
\frac{\partial {\mathcal F}^{}_G}{\partial a_\gamma}
+
\sum_{\alpha=c,q}
\frac{\partial v^{}_{\alpha}}{\partial a_\gamma}
\frac{\partial {\mathcal F}^{}_G}{\partial v^{}_{\alpha}}
=
\frac{\partial {\mathcal F}^{}_G}{\partial a_\gamma} \, ,
\label{1std}
\end{align}
%----------------------------------------------------------
where we have used Eq.~(\ref{eq:saddlepoint1}). 
To obtain the second derivative, we completely differentiate  Eq. (\ref{1std})   
%----------------------------------------------------------
\begin{align}
&\frac{d^2 {\cal F}}{d a^{}_{\gamma'} \, d a^{}_\gamma}
=
\frac{\partial^2 {\mathcal F}^{}_G}{\partial a^{}_{\gamma'} \, \partial a^{}_{\gamma}}
+
\sum_{\alpha=c,q}
\frac{d v_{\alpha}}{d a_{\gamma'}}
\frac{\partial^2 {\mathcal F}_G}{\partial a_{\gamma}^{}\, \partial v^{}_{\alpha}}
\nonumber\\
&=
\frac{\partial^2 {\mathcal F}_G}{\partial a_{\gamma'}^{}\, \partial a_{\gamma}^{}}
+
\sum_{\alpha,\alpha'=c,q}
\frac{\partial^2 {\mathcal F}^{}_G}{\partial a_{\gamma'}^{}\, \partial v^{}_{\alpha}}
\,
U^{}_{\alpha \alpha'}
\,
\frac{\partial^2 {\mathcal F}^{}_G}{\partial a_{\gamma}^{} \, \partial v^{}_{\alpha'}} \, , 
\label{2ndd}
\end{align}
%----------------------------------------------------------
where we have used Eq. (\ref{evc}). 
Note that the final  form is symmetric in $a_\gamma$ and $a_{\gamma '}$. 
The third derivative can be obtained by further differentiating Eq.~(\ref{2ndd}). 
The lengthy expression of the third derivative is relegated to Appendix \ref{3d}.

\begin{figure}[ht]
\includegraphics[width=0.9 \columnwidth]{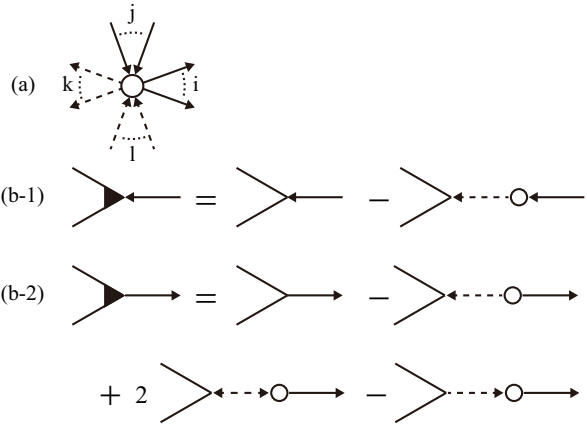}
\caption{
(a) A ``bare" vertex (empty circle) $L^{ik}_{j \ell}$; (b-1) and (b-2) corrected vertices (solid triangles). 
}
\label{vc}
\end{figure}

The derivatives of the CGF of the two-terminal junction given in  Eqs.~(\ref{1std}), (\ref{2ndd}),  and (\ref{3rdd}) indicate the explicit form of the vertex corrections: the full derivative is obtained upon inserting  
%----------------------------------------------------------
\begin{align}
\frac{d}{d a_{\gamma}^{}}
\to 
\frac{\partial}{\partial a_{\gamma}^{}}
+
\sum_{\alpha'=c,q}
\frac{\partial^2 {\mathcal F}^{}_G }{\partial v^{}_{\alpha'} \partial a_{\gamma}^{}}
U^{}_{\alpha' \alpha}
\frac{\partial}{\partial v^{}_\alpha}
\, 
\label{gevc}
\end{align}
%----------------------------------------------------------
into each  bare vertex. Let us now explore these vertex corrections in detail. 
Note that in  deriving the transport coefficients one has to set $\lambda={\mathcal A}=0$ after performing the differentiations [see   Eq.~(\ref{nontracoecha})]. Under these circumstances the saddle-point solution is  $\lambda_P^*={\mathcal A}_P^*=0$, and consequently
the matrix $U$, Eq. (\ref{u}), becomes
%----------------------------------------------------------
\begin{align}
U
&
\to 
-
\left.
\left[
\begin{array}{cc}
\frac{\partial^2 {\mathcal F}^{}_G }{\partial {\mathcal A}^{}_P \partial {\mathcal A}_P^{}}  &  \ 
\frac{\partial^2 {\mathcal F}^{}_G }{\partial {\mathcal A}_P^{} \partial (i \lambda_P^{}) }\\
\frac{\partial^2 {\mathcal F}^{}_G }{\partial (i \lambda^{}_P) \partial {\mathcal A}^{}_P} &  \ 
\frac{\partial^2 {\mathcal F}^{}_G }{\partial (i \lambda^{}_P) \partial (i \lambda^{}_P)}
\end{array}
\right]^{-1}
\right|_{\lambda=\lambda_P={\mathcal A}={\mathcal A}_P=0}
\nonumber \\
&=
\frac{1}{L^{0 \, 1}_{0 \, 1, +}}
\left[
\begin{array}{cc}
\ 2 & -1 \\ & \\
-1 &\  0
\end{array}
\right]
\, . 
\end{align}
%----------------------------------------------------------
The last equality here is obtained upon using Eqs. (\ref{1eq3}) and (\ref{2eq3}). 
Similarly, the matrix $\partial^2 {\mathcal F}_G/\partial v_k \partial a_i$ is 
%----------------------------------------------------------
\begin{align}
\left.
\left[ \! 
\begin{array}{cc}
\frac{\partial^2 {\mathcal F}^{}_G }{\partial {\mathcal A}^{}_P \partial {\mathcal A} } &  
\frac{\partial^2 {\mathcal F}^{}_G }{\partial {\mathcal A}^{}_P \partial (i \lambda) } \\
\frac{\partial^2 {\mathcal F}^{}_G }{\partial (i \lambda^{}_P) \partial {\mathcal A}} &  
\frac{\partial^2 {\mathcal F}^{}_G }{\partial (i \lambda^{}_P) \partial (i \lambda)}
\end{array}
\! \right]
\right|_{\lambda=\lambda_P={\mathcal A}={\mathcal A}_P=0}
=
\left[\! 
\begin{array}{cc}
0 &  {L}^{10}_{01} \\  &  \\
{L}^{01}_{10} &  {L}^{11}_{00} \\
\end{array}
\! \right]
\, . 
\nonumber \\
\end{align}
%----------------------------------------------------------
Collecting these results, Eq. (\ref{gevc}) can be written explicitly as
(for $\lambda=\lambda_P={\mathcal A}={\mathcal A}_P=0$)
%----------------------------------------------------------
\begin{align}
\frac{d}{d {\mathcal A}}
& \to 
\frac{\partial}{\partial {\mathcal A}}
-
\frac{{L}^{01}_{10}}{L^{01}_{01, +}}
\frac{\partial}{\partial {\mathcal A}^{}_{P}}
\, ,
%\label{vc1}
\nonumber\\
\frac{d}{d {\mathcal (i \lambda) }}
& \to 
\frac{\partial}{\partial (i \lambda)}
+
\frac{
2 L^{10}_{01} - L^{11}_{00}
}{L^{01}_{01,+}}
\frac{\partial}{\partial {\mathcal A}^{}_{P}}
-
\frac{L^{10}_{01}}{L^{01}_{01,+}}
\frac{\partial}{\partial (i \lambda^{}_{P})}
\, . 
\label{vc2}
\end{align}
%----------------------------------------------------------
Following Ref. \onlinecite{Nagaev1} the vertex corrections 
can be visualized diagrammatically. Figure \ref{vc}(a) depicts a ``bare" vertex, connecting 
$i$ outgoing solid lines, $k$ outgoing dotted lines, $j$ incoming solid lines,  and $\ell$ incoming dotted lines which correspond to 
${L}^{ik}_{j \ell}$. 
The dressed vortices  presented in Eqs. (\ref{vc2}) are shown in Figs. \ref{vc}(b-1) and \ref{vc}(b-2), 
respectively.

%----------------------------------------------------------
%----------------------------------------------------------

\subsection{Linear and lowest nonlinear transport coefficients}
\label{vercor}

Here we exploit the general formulae derived in Sec. \ref{subvc} to present expressions for  the  transport coefficients of the heat current in the linear-response regime, and the lowest nonlinear ones.

\begin{figure}[ht]
\includegraphics[width=0.95 \columnwidth]{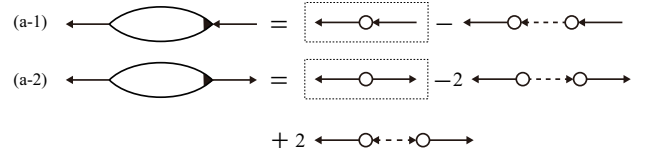}
\caption{
Diagrams for the  linear-response conductance (a-1) and the equilibrium noise (a-2). 
Only the ``bare" diagrams, indicated by dotted squares, are included in
the minimal-correlation coordinate approach explained in Appendix \ref{mcc}.}
\label{diagnoise}
\end{figure}

Figures~\ref{diagnoise} (a-1) and (a-2) portray the diagrams
for the linear-response conductance and the equilibrium noise, respectively, leading to the expressions [see Eq. (\ref{nontracoecha})]
%----------------------------------------------------------

\begin{align}
L^{2}_{0}/2
=
L^{1}_{1}
=&
L^{1 \, 0}_{1 \, 0 , +}
-
\frac{ (L^{1 \, 0}_{0 \, 1 , +} )^{2}_{} - (L^{0 \, 1}_{1 \, 0, -} )^{2}_{} }
{L^{0 \, 1}_{0 \, 1 , +}}
\, , 
\nonumber\\
L^{1 \, 0}_{1 \, 0 , -}
=&
0
\, , \label{lrn}
\end{align}
%----------------------------------------------------------
which satisfy the fluctuation-dissipation theorem and the Onsager relations [see Eqs. (\ref{1eq1t}) and (\ref{1eq3t}) and the discussion around them]. 
Furthermore, by  Eqs.~(\ref{conlaw}) and (\ref{trans}), Eqs. (\ref{lrn}) reproduce the  well-known result (see, e.g.,  Ref.~\onlinecite{Buettiker}  for the electric-transport analogue), 
%----------------------------------------------------------
\begin{align}
L^{1}_{1}
=
K_{LL}
-
\frac{ K_{LP} K_{PL} }{K_{PP}}
\, \ , 
\end{align}
%----------------------------------------------------------
where $K_{ab}$ are the elements of the three-terminal heat-conductance matrix defined in Eq. (\ref{linheacon}).

%----------------------------------------------------------
%----------------------------------------------------------

\begin{figure}[ht]
\includegraphics[width=1 \columnwidth]{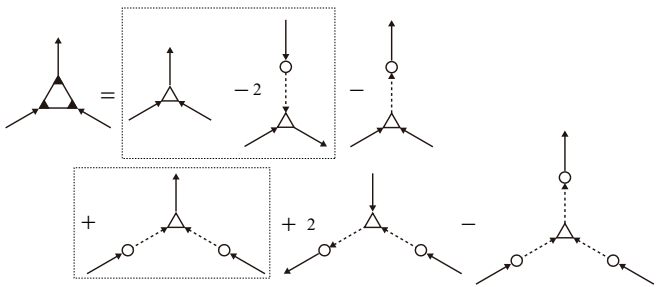}
\caption{
The six diagrams comprising   the second nonlinear conductance. The three diagrams enclosed in the dotted squares are the ``bare" one and its cascade corrections, which are accounted for in the minimal-correlation coordinate approach (Appendix \ref{mcc}). 
The second and fifth diagrams on the right-hand side contain 
$L^{10}_{11}$
and 
$L^{01}_{11}$,
respectively, and thus vanish. 
}
\label{diagg2}
\end{figure}

There are six vertex-correction diagrams comprising the lowest nonlinear conductance; these are shown in Fig. ~\ref{diagg2}, leading to the expression 
%----------------------------------------------------------
\begin{align}
&L^{1}_{2}
=
L^{1 \, 0}_{2 \, 0}
-
\frac{
L^{1 \, 0}_{0 \, 1} L^{0 \, 1}_{2 \, 0}
+
2 
L^{0 \, 1}_{1 \, 0} L^{1 \, 0}_{1 \, 1}
}
{L^{0 \, 1}_{0 \, 1}}
\nonumber \\
&+
\frac{ 
(L^{0 \, 1}_{1 \, 0})^2 L^{1 \, 0}_{0 \, 2}
+
2 
L^{0 \, 1}_{1 \, 0} L^{1 \, 0}_{0 \, 1} L^{0 \, 1}_{1 \, 1}
}{ (L^{0 \, 1}_{0 \, 1})^2 }
-
\frac{
(L^{0 \, 1}_{1 \, 0})^2 L^{1 \, 0}_{0 \, 1} L^{0 \, 1}_{0 \, 2}
}
{(L^{0 \, 1}_{0 \, 1})^3}
\label{G2full}
\, . 
\end{align}
%----------------------------------------------------------
One notes  the appearance of the second nonlinear thermal conductances of the three-terminal junction, 
$L^{1 \, 0}_{1 \, 1}$
$L^{1 \, 0}_{0 \, 2}$, 
$L^{0 \, 1}_{2 \, 0}$, 
$L^{0 \, 1}_{1 \, 1}$, 
and 
$L^{0 \, 1}_{0 \, 2}$, which enter this expression 
because of the changes in the temperature of the probe electrode. However, as can be seen from Eq. (\ref{cgfsm}), the simultaneous derivative of the 
CGF with respect to  ${\mathcal A}$ and ${\mathcal A}_P$ vanish, 
%----------------------------------------------------------
\begin{align}
L^{1 \, 0}_{1 \, 1}
=
L^{0 \, 1}_{1 \, 1}
=
0
\label{simres}
\, , 
\end{align}
%----------------------------------------------------------
and therefore Eq. (\ref{G2full}) is simplified, 
%----------------------------------------------------------
\begin{align}
&L^{1}_{2}
=
L^{1 \, 0}_{2 \, 0}
-
\frac{
L^{1 \, 0}_{0 \, 1} L^{0 \, 1}_{2 \, 0}
}
{L^{0 \, 1}_{0 \, 1}}
+
\frac{ 
(L^{0 \, 1}_{1 \, 0})^2 L^{1 \, 0}_{0 \, 2}
}{ (L^{0 \, 1}_{0 \, 1})^2 }
-
\frac{
(L^{0 \, 1}_{1 \, 0})^2 L^{1 \, 0}_{0 \, 1} L^{0 \, 1}_{0 \, 2}
}
{(L^{0 \, 1}_{0 \, 1})^3}
\label{G2fulln}
\, . 
\end{align}
%----------------------------------------------------------

%----------------------------------------------------------
%----------------------------------------------------------

\begin{figure}[ht]
\includegraphics[width=1 \columnwidth]{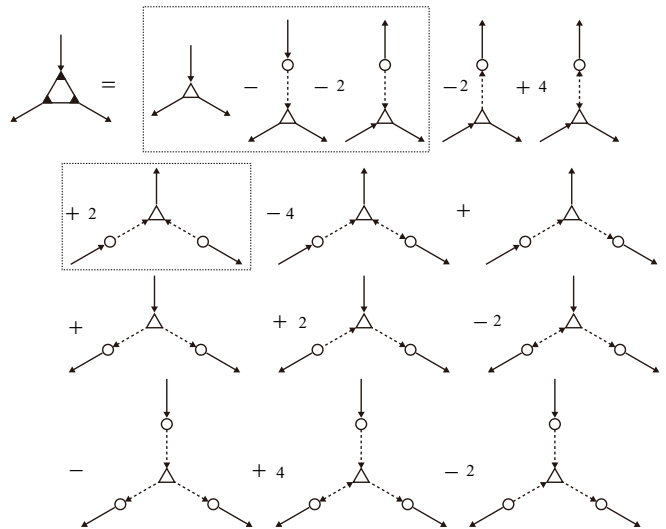}
\caption{
The fourteen diagrams contributing to the expression for the linear-response  noise. The four diagrams enclosed in the dotted squares  are accounted for in the minimal-correlation coordinate approach (Appendix \ref{mcc}). 
The third and fifth diagrams on the right-hand side contain 
$L^{10}_{11}$
and  the 10th and the 11th ones contain 
$L^{01}_{11}$. These diagrams therefore vanish.
}
\label{diags1}
\end{figure}

The expression for the linear-response  noise results from the diagrams  depicted in Fig.~\ref{diags1} and turns out to be rather complicated even after utilizing  Eq.~(\ref{simres}). 
However, by using the universal relations among the bare transport coefficients of the three-terminal junction, Eqs.~(\ref{eq6a}), 
it is possible to show that the components symmetric with respect to a magnetic field and the anti-symmetric ones obey
%----------------------------------------------------------
\begin{align}
L^1_{2,+}
&=
L^2_{1,+}
\, ,
\nonumber
\\
L^1_{2,-}
&=
L^2_{1,-}/3
\, . 
\label{uninon}
\end{align}
%----------------------------------------------------------
[See Eqs.~(\ref{magsym}) and (\ref{magasym}) for the explicit expressions.] 
These relations between the noise in linear-response regime and the second nonlinear conductance, agree with the universal relations derived in Ref. \onlinecite {SU}  [see Eqs.~(\ref{eq1at})].

%----------------------------------------------------------
%----------------------------------------------------------

The number of the diagrams is reduced considerably when one adopts the approach of the minimal-correlation coordinate, \cite{Jordan} see Appendix \ref{mcc} and in particular  Eq. (\ref{mincor}).  Then, only 3 (4) diagrams remain for the second nonlinear conductance (the linear-response noise)   as indicated by dotted squares in Fig. \ref{diagg2} (Fig. \ref{diags1}). 
This is obviously advantageous for a practical use, e.g. for a numerical calculation. 
However, since the bare three-terminal transport coefficients do not satisfy the universal relations presented in Appendix~\ref{nonlineonsager}, 
it is not possible to prove the symmetry Eq.~(\ref{uninon}) analytically [it is possible for our case as given in Eqs.~(\ref{magsym}) and (\ref{magasym})], 
which is, in our opinion,  a weak point.

\subsection{Three-terminal triple-quantum dot}
\label{tp}

\begin{figure}[ht]
\includegraphics[width=.9 \columnwidth]{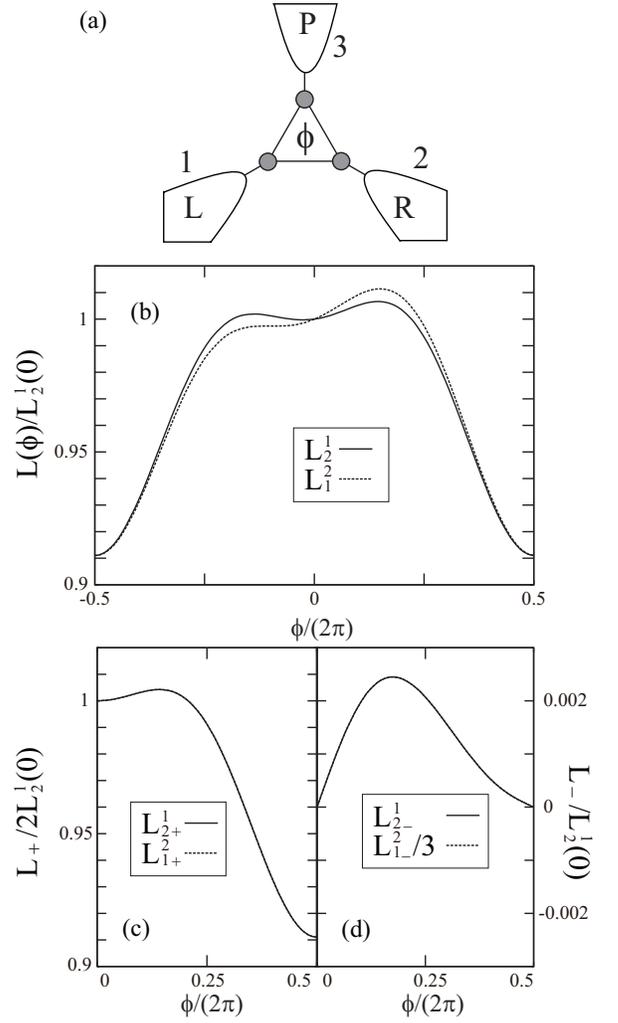}
\caption{
(a) Schematic picture of three-terminal triple quantum dot. 
The flux threads  the triangular region. 
(b) The second  nonlinear thermal conductance (solid line) and 
the linear-response  thermal noise (dotted line) as a function of the flux. 
$L^1_2(0)=1.39 \times 10^{-4} \Gamma^3/(e^2 R_{K})$, 
where $R_K=h/e^2$ is the resistance quantum. 
Panels (c) and (d) show the symmetric (solid lines) and anti-symmetric (dashed lines) components in the flux. 
Solid and dashed lines in each panel overlap each other.
Parameters: 
$t=0.1 \Gamma$, 
$\epsilon_0=-0.5 \Gamma$, 
$\beta \Gamma=10$,
and
$x=0.25$.
}
\label{g2s1}
\end{figure}

The general results presented in the previous sections are explicitly illustrated in this section by analyzing 
a
triple quantum-dot connected to three terminals and threaded by a
magnetic flux $\Phi=Bs$, as shown schematically 
in Fig.~\ref{g2s1}(a) ($s$ is the relevant area). 
The system is described by the Hamiltonian
%----------------------------------------------------------
\begin{align}
{\cal H} & =
\sum_{j=1}^3 \epsilon^{}_0 d_j^\dagger d^{}_j
+
\sum_{j=1}^{3}
\sum_k
\epsilon^{}_{k}a^{\dagger}_{jk}a^{}_{jk}
\nonumber\\
&
+
\sum_{j=1}^3 
\left(
t {\rm e}^{i \phi/3} 
d_{j+1}^\dagger d^{}_j 
+
\sum_k
t^{}_{jk} d_j^\dagger a^{}_{j k} 
+{\rm H.c}
\right) 
\,\  ,\label{H}
\end{align}
where for simplicity the spin degree of freedom is ignored. 
The first term in Eq. (\ref{H}) pertains to the three uncoupled dots (with each dot represented  by a single energy level), the second describes the three electrodes (assuming each to consist of a free electron gas. 
To make the expression compact we identify the left, right and probe electrodes with the 1st, 2nd and 3rd electrodes, respectively) and the third gives the tunneling between neighboring dots and between  each dot and the electrode to which it is attached. 
In Eq. (\ref{H}),  $d_j$ annihilates  an electron on the $j$-th dot  (with
$d_4\equiv d_1$), 
$a_{jk}$ destroys an electron with  wave vector $k$ in the $j$-th electrode, 
$t$ is the hopping matrix element between adjacent  dots,  and $t_{jk}$ is the tunneling matrix element between the
$j$-th dot  and $j$-th electrode. 
The effect of a magnetic field is contained in the Aharonov-Bohm phase        $\phi=\Phi/\Phi_0$, where 
$\Phi_0=\hbar c/e$ is the flux quantum; hence
$\phi$ reverses its sign when
$B \to -B$.

The scattering matrix of the triple quantum-dot system comprises  the following elements
%----------------------------------------------------------
\begin{align}
&S_{j+1 \, j}
(\omega;B)
=
\{
4 {\rm e}^{-i \phi/3}
t \Gamma [\Gamma - 2i (\omega-\epsilon_0 + t {\rm e}^{i \phi})]
\}
/\Delta(\phi)
\, ,
\nonumber \\
&S_{j \, j+1}(\omega;B)
=
S_{j+1 \, j}(\omega;-B)
\, ,
\nonumber\\
&S_{jj}(\omega;B)
=
\{ 
[2 (\omega-\epsilon_0) - i \Gamma]
[2 (\omega-\epsilon_0) + i \Gamma]^2
\nonumber \\
&
- 4 \, t^2
[6 (\omega-\epsilon_0) + i \Gamma]
-
16 \, t^3 \cos \phi 
\} / \Delta(\phi)
\, ,
\end{align}
where
\begin{align}
\Delta(\phi)
&=
[2 (\omega-\epsilon_0) + i \Gamma]^3 
-12 \, t^2 
[2 (\omega-\epsilon_0) + i \Gamma]
\nonumber \\
&
-
16 \, t^3 \cos \phi 
\, . 
\end{align}
%----------------------------------------------------------
We have  assumed here that the tunnel coupling strength is energy independent, and introduced the level broadening
%----------------------------------------------------------
\begin{align}
\Gamma
=
2 \pi \sum_k |t_{jk}|^2 \delta (\omega-\epsilon_k)
\, .
\end{align}
%----------------------------------------------------------

Figure~\ref{g2s1}(b) shows the flux dependence of the second nonlinear thermal conductance $L^1_2$ (solid line) and the liner-response expression for the heat-current noise $L^2_1$ (dotted line). 
As can be seen, the curves are not symmetric with respect to the flux $\phi$. 
Such an anti-symmetric component induced by the magnetic field at out-of-equilibrium conditions is absent for the  two-terminal conductor of noninteracting electrons, but is finite for the setup with a probe 
(for the magnetic field induced electric heat current asymmetry, see e.g. Ref. \onlinecite{Bedkihal}). 
From Fig.~\ref{g2s1}(b), we may conclude that $L^1_2$ (solid line) and $L^2_1$ (dotted line) are uncorrelated. 
However, when we look at the symmetrized and the anti-symmetrized components [defined in Eq.~(\ref{sym})] the correlation becomes clearer. 
Figure~\ref{g2s1}(c) shows the symmetrized components, $L^1_{2,+}$ (solid line) and $L^2_{1,+}$ (dotted line). 
Figure~\ref{g2s1}(d) shows the anti-symmetrized components (with 1/3 for the linear response of noise), $L^1_{2,-}$ (solid line) and $L^2_{1,-}/3$ (dotted line). 
The two panels show that the solid lines and the dotted lines overlap, 
which means that the FT Eq. (\ref{uninon}) is satisfied. 

%----------------------------------------------------------
%----------------------------------------------------------

A recent study of the Langevin equation with non-Gaussian white noise suggests that non-Gaussian corrections around the saddle point play a  crucial role in certain cases. \cite{Kanazawa}  
We have therefore checked numerically that the saddle-point solution provides a physically reasonable probability distribution in the case of the triple dot. 
In the limit of  long measurement times $\tau$, the inverse Fourier transform of the characteristic function can be written approximately as a Legendre-Fenchel transform of the scaled CGF (\ref{fscgf}), 
%----------------------------------------------------------
\begin{align}
&P_\tau
(
\{ \epsilon_r \}, \{ {\mathcal A}_r \};B 
)
\nonumber\\
&=
\frac{1}{2\pi}
\int 
d \lambda
e^{
-i {\lambda} \varepsilon_L
+
\tau 
{\mathcal F}
(\lambda,
{\mathcal A}
;B,x
)
}
\delta(\varepsilon_L + \varepsilon_R)
\nonumber\\
&\approx 
e^{-\tau {\mathcal I}}
\, 
\delta(\varepsilon_L + \varepsilon_R)
\, , 
\label{pdldf}
\end{align}
%----------------------------------------------------------
where $\epsilon_{L}$ and $\epsilon_R$ are electron energies flowing out of the left(1st) electrode and the right(2nd) electrode, respectively. 
The cumulant generating function ${\cal F}$ was given in Eq.~(\ref{sadapp}) 
with the $S$-matrix for the three-terminal triple-quantum dot. 
${\mathcal I}$  is the rate function~\cite{Touchette}
%----------------------------------------------------------
\begin{align}
{\mathcal I}
=
\max_\lambda
\left[
i \lambda I_E
-
{\mathcal F}
(\lambda,
{\mathcal A}, 
;B,x
)
\right]
\, ,
\end{align}
%----------------------------------------------------------
with $\lambda$ being a purely  imaginary number. 
This result verifies that the heat current flowing out of the left junction and that flowing into the right junction are identical,  
%----------------------------------------------------------
$
I_E =
\lim_{\tau \to \infty}
\varepsilon_L/\tau 
=
\lim_{\tau \to \infty}
-\varepsilon_R/\tau 
$. 
%----------------------------------------------------------

\begin{figure}[ht]
\includegraphics[width=.9 \columnwidth]{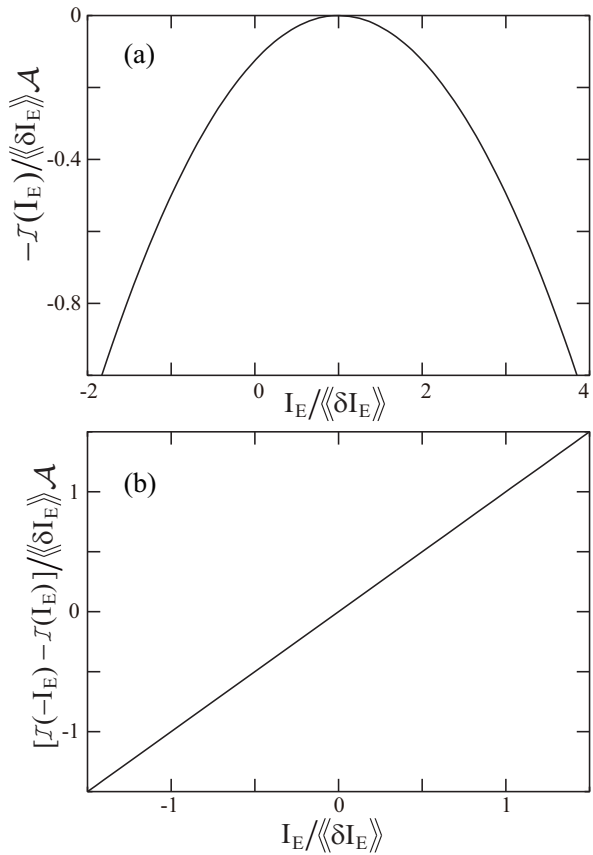}
\caption{
(a) The rate function and (b) the fluctuation theorem for $\phi=0$. 
Parameters: 
${\mathcal A} \Gamma=0.5$, 
$\beta \Gamma=10$. 
}
\label{rfft}
\end{figure}

Figure \ref{rfft}(a) depicts the rate function,  calculated by numerically solving the saddle-point equations (\ref{sadeqsd}). 
This procedure yields a single solution and it results in a reasonable rate function as shown in Fig.~\ref{rfft}(a). 
Furthermore, by using Eqs. (\ref{ftg}) and (\ref{pdldf}), the FT in the limit of long measurement times can be expressed as 
${\mathcal I}(-I_E) - {\mathcal I}(I_E) = I_E {\mathcal A}$. 
%----------------------------------------------------------
This equality is plotted in Fig.~\ref{rfft}(b), which further support  the  validity of our numerical solution.

%----------------------------------------------------------
%----------------------------------------------------------

\section{Summary}
\label{summary}

We have  investigated the fluctuation theorem of the heat transfer driven by  a  temperature difference across a three-terminal conductor for which one of the electrodes serves as a probe electrode used to measure the conductor temperature.  
The stochastic energy fluctuations in the probe electrode are integrated over to yield the cumulant-generating function of the reduced two-terminal junction.
We have proven that this 
generating function satisfies the two-terminal fluctuation theorem.
We have obtained expressions for the second nonlinear conductance and  the linear-response noise, and have shown explicitly that they obey the symmetries 
imposed by the FT. Furthermore, we have shown that in order for this symmetries to hold, it is imperative to account for all vertex corrections entering the expressions for the transport coefficients. 
We  stress that our expressions for the transport coefficients {\it explicitly} preserve the symmetry of the fluctuation theorem [Eqs.~(\ref{magsym}) and (\ref{magasym})], which is not the case for those drawn from the approach of the minimal-correlation coordinate. \cite{Jordan} 
We have applied our theory to a three-terminal triple quantum-dot system. 

We have included in our analysis solely the thermodynamic driving forces resulting from temperature differences. 
A full treatment of thermoelectricity requires the inclusion of chemical-potential gradients, as well as voltage probes; these will be handled in  future publications.

\section*{Acknowledgments} 

We thank Dimitri Golubev, Hisao Hayakawa and Kiyoshi Kanazawa for valuable discussions. 
This work was supported by the Grant-in-Aid for Young Scientists (B) (No. 23740294) and MEXT kakenhi ``Quantum Cybernetics" (No. 21102003), 
the Israeli Science Foundation (ISF), and the US-Israel Binational Science Foundation (BSF).
O. Entin-Wohlman and A. Aharony acknowledge gratefully the hospitality of Tsukuba University, where this work was completed. 
\appendix
%\begin{appendix}

\begin{widetext}
\section{Proof of the FT for the CGF in terms of the scattering matrix}
\label{ftsm}

In order to keep the paper self-contained, 
we outline in this appendix the proof that the CGF as given in terms of the scattering matrix ${\mathbf S}$  [see Eq. (\ref{cgfsm})] obeys the symmetries derived from the FT, 
i.e., Eq. (\ref{ftcgfo}).
To this end we adopt the representation of the 
scattering matrix introduced  in   Refs. \onlinecite{Levitov} and \onlinecite{FB2} (see Appendix A in the latter). 
For convenience of notations, we denote here the 
left ($r=L$), the right ($r=R$), and the probe ($r=P$) electrodes by  1, 2,  and 3, respectively. 

The first step is to decompose the determinant in the integrand of Eq. (\ref{cgfsm})
into  terms describing  multi-particle scattering events by using the Cauchy-Binet formula, 
%----------------------------------------------------------
\begin{align}
\det
[{\mathbf 1}-{\mathbf f} (\omega){\mathbf K}(\mat {\lambda},\omega;B)]
&=
\sum_{ I }
\sum_{ O }
\left|
{\mathbf S}(\omega;B)^{ O }_{ I }
\right|^2
\Big (
\prod_{r \in I , r' \not \in I }
f_r^{} 
(1-f^{}_{r'})
\Big )
\exp \! 
\Big ( -i  \sum_{s \in I} \lambda_s \omega 
+i \sum_{s' \in O} \lambda_{s'} \omega 
\Big )
\nonumber\\
&=
\sum_{ I }
\sum_{ O }
\left|
{\mathbf S}(\omega;B)^{ O }_{ I }
\right|^2
\Big (
\prod_{r}
(1-f^{}_r)
\Big )
\exp \! 
\Big ( -i  \sum_{s \in I } (\lambda^{}_s - i \beta^{}_s) \omega 
+i \sum_{s' \in O} \lambda^{}_{s'} \omega 
\Big )
\, .\label{decom}
\end{align}
%----------------------------------------------------------
Here $I$ is a set of terminals from which particles are emerging, 
and $O$ is the one in which they are being absorbed. 
For a three-terminal junction, 
$I, O=\{1,2,3 \}$ 
or 
$I, O=\{1,2 \}, \{ 2,3 \}, \{ 1,3 \}$ 
or 
$I, O=\{1 \}, \{ 2 \}, \{ 3 \}$. The square matrix
${\mathbf S}^{ O }_{ I }$ is %the square matrix, which is 
one of submatrices of the scattering  matrix ${\mathbf S}$ with rows $O$ and columns $I$, and  
$\left|{\mathbf S}^{ O }_{ I }\right|$ is its determinant, i.e., it is the minor. 
For example,  $I=\{2,3 \}$ and $O=\{1,2 \}$ means that particles are emerging  from  terminals $R$ and $P$ and then absorbed in terminals $L$ and $R$. 
The submatrix is then 
%----------------------------------------------------------
\begin{align}
{\mathbf S}^{ \{1,2 \} }_{ \{2,3 \} }
=
\left[
\begin{array}{cc}
S_{LR} & S_{LP} \\
S_{RR} & S_{RP} 
\end{array}
\right]
\, .
\end{align}
%----------------------------------------------------------
Next, the transformation 
$ \lambda_r \to -\lambda_r + i {\mathcal A}_r $
is applied to Eq. (\ref{decom}) to yield 
%----------------------------------------------------------
\begin{align}
\det
[{\mathbf 1}-{\mathbf f} (\omega){\mathbf K}(\mat {\lambda},\omega;B)]
&\to
\sum_{ I}
\sum_{ O}
\left|
{\mathbf S}(\omega;B)^{ O }_{ I }
\right|^2
\Big (
\prod_{r}
(1-f^{}_r)
\Big )
\exp \! 
\Big ( i  \sum_{s \in I} \lambda^{}_s \omega 
-i \sum_{s' \in O} (\lambda^{}_{s'} - i \beta^{}_{s'}) \omega 
\Big )\nonumber\\
&=
\sum_{I}
\sum_{O}
\left|
{\mathbf S}(\omega;-B)^O_I
\right|^2
\Big (
\prod_{r}
(1-f^{}_r)
\Big )
\exp \! 
\Big ( 
-i \sum_{s' \in O} (\lambda^{}_{s'} - i \beta^{}_{s'}) \omega 
+
i  \sum_{s \in I} \lambda^{}_s \omega 
\Big )
\nonumber\\
&
=
\det
[{\mathbf 1}-{\mathbf f}  (\omega){\mathbf K}(\mat {\lambda},\omega;-B)]
\, , \label{pro}
\end{align}
%----------------------------------------------------------
where we have used the micro-reversibility condition
Eq. (\ref{micrev}). Equation (\ref{pro}) proves that the CGF in its representation  
(\ref{cgfsm}) (valid for noninteracting electrons) obeys the FT.

\end{widetext}

\section{Relations among the transport coefficients}
\label{nonlineonsager}

The fluctuation theorem imposes certain relations among the transport coefficients.~\cite{SU}
Here we summarize several of them. 
In order to derive those one considers the symmetrized and anti-symmetrized forms of the FT as applied to the two-terminal CGF, 
%----------------------------------------------------------
\begin{align}
&
{\mathcal F}(\lambda,{\mathcal A};B,x)
\pm
{\mathcal F}(\lambda,{\mathcal A};-B,x)
\nonumber \\
&
=
{\mathcal F}(-\lambda+i {\mathcal A},{\mathcal A};-B,x)
\pm
{\mathcal F}(-\lambda+i {\mathcal A},{\mathcal A};B,x)
\, , \label{SAS}
\end{align}
%----------------------------------------------------------
and the symmetrized and anti-symmetrized coefficients 
\begin{align}
L^{n}_{m,\pm}(B)=L^{n}_{m}(B)\pm L^{n}_{m}(-B)\ ,
\end{align}
where $L^{n}_{m}$ is defined in Eq. (\ref{nontracoecha}). 
By expanding both sides of Eq. (\ref{SAS})
in powers of the counting field $\lambda$ and the affinity ${\mathcal A}$
and comparing the resulting coefficients, it is found that
%----------------------------------------------------------
\begin{align}
L^{n}_{m \pm}
&=
\pm
\sum_{k=0}^{m}
\left( \! 
\begin{array}{c}
m \\
k
\end{array}
\! \right)
(-1)^{k+n}
L^{n+k}_{m-k \,\pm}
\, . 
\label{2tftc1}
\end{align}
%----------------------------------------------------------
The normalization of the CGF implies that 
${\mathcal F}(0,{\mathcal A};B)=0$, 
and consequently
%----------------------------------------------------------
\begin{align}
L^{0}_{n}=0 \, .
\label{2tftc2}
\end{align}
%----------------------------------------------------------

The lowest nontrivial coefficient, for which $m=n=1$, is [see Eq. (\ref{2tftc1})]
%----------------------------------------------------------
\begin{align}
L^{1}_{1 \pm}
&=
\pm
\left(
-L^{1}_{1 \,\pm}
+L^{2}_{0 \,\pm}
\right)
\, . 
\end{align}
%----------------------------------------------------------
This relation yields that the linear-response coefficients obey 
the fluctuation-dissipation theorem relating the linear-response thermal conductance to its noise,  
%----------------------------------------------------------
%\begin{align}
\begin{align}
L^{1}_{1+}
\! &= \!
L^{2}_{0+}/2
\, ,
\label{1eq1t}
\end{align}
and the Onsager relation imposing that the linear-response thermal conductance and also its noise are even in the magnetic field, 
\begin{align}
L^{1}_{1 -}
\! &= \!
L^{2}_{0 -}
=0
\, . 
\label{1eq3t}
%\end{align}
\end{align}
%----------------------------------------------------------
Similarly, Eqs. (\ref{2tftc1}) and (\ref{2tftc2}) produce interrelations among the coefficients which hold beyond linear response, e.g., 
%----------------------------------------------------------
\begin{align}
L^{1}_{2+} &= L^{2}_{1+} ,
%\label{eq2t}
\nonumber\\
L^{1}_{2-} &= L^{2}_{1-}/3=L^{3}_{0-}/6 ,
%\label{eq3t}
\nonumber\\
L^{3}_{0 +} &= 0 \, .
\label{eq1at}
\end{align}
%----------------------------------------------------------

For completeness, we present below similar relations  for a three-terminal junction. 
By using Eq.~(\ref{ftfg}), these are determined by an expression analogous to Eq. (\ref{2tftc1}), 
%----------------------------------------------------------
\begin{align}
L^{n_1 \, n_2}_{m_1 \, m_2 \pm}
&=
\pm
\sum_{k_1=0}^{m_1}
\sum_{k_2=0}^{m_2}
\left( \! 
\begin{array}{c}
m_1 \\
k_1
\end{array}
\! \right)
\left( \! 
\begin{array}{c}
m_2 \\
k_2
\end{array}
\! \right)
\nonumber \\
& \times
(-1)^{k_1+n_1+k_2+n_2}
L^{n_1+k_1 \, n_2+k_2}_{m_1-k_1 \, m_2-k_2 \pm}
\, . 
\end{align}
%----------------------------------------------------------
From this expression, in conjunction with the normalization condition
$L^{0 \, 0}_{m_1 \, m_2 \pm}=0$, 
one can obtain several classes of interrelations. % among the  transport coefficients. 
First, there are the ones of the linear-response regime, which obey the fluctuation-dissipation theorem
%----------------------------------------------------------
\begin{align}
L^{1 \, 0}_{1 \, 0, +}
&=
L^{2 \, 0}_{0 \, 0, +}/2
\, ,\nonumber\\
%\label{1eq1}
L^{0 \, 1}_{0 \, 1, +}
&=
L^{0 \, 2}_{0 \, 0, +}/2
\, .
\label{1eq3}
\end{align}
%----------------------------------------------------------
Second, there are the Onsager-Casimir relations, 
%----------------------------------------------------------
\begin{align}
L^{1 \, 0}_{0 \, 1 , +}
&=
L^{0 \, 1}_{1 \, 0 , +}
=
L^{1 \, 1}_{0 \, 0 , +}/2
%\label{2eq1}
\, ,\nonumber
\\
L^{1 \, 1}_{0 \, 0 , -}
&=
0
\, ,\nonumber
%\label{2eq2}
\\
L^{1 \, 0}_{0 \, 1 , -}
&=
-
L^{0 \, 1}_{1 \, 0 , -}
\, ,
\nonumber\\
%\label{1eq2}
L^{1 \, 0}_{1 \, 0, -}
&=
L^{2 \, 0}_{0 \, 0, -}
=
L^{0 \, 1}_{0 \, 1 , -}
=
L^{0 \, 2}_{0 \, 0, -}
=0
\, . 
\label{2eq3}
\end{align}
%----------------------------------------------------------
These symmetries are identical to  those obtained for electrical transport. \cite{Buettiker} 
Third, there are relations for the higher transport coefficients, beyond the linear-response regime, 
%----------------------------------------------------------
\begin{align}
L^{3\,0}_{0\,0,+} &= 0 \ ,\nonumber
%\label{eq1}
\\
L^{1\,0}_{2\,0,+} &= L^{2\,0}_{1\,0,+} \ ,\nonumber
%\label{eq2}
\\
L^{1\,0}_{2\,0,-} &= L^{2\,0}_{1\,0,-}/3=L^{3\,0}_{0\,0,-}/6\ ,\nonumber
%\label{eq3}
\\
L^{0\,3}_{0\,0,+} &= 0 \, ,\nonumber
%\label{eq1a}
\\
L^{0\,1}_{0\,2,+} &= L^{0\,2}_{0\,1,+} \, ,\nonumber
%\label{eq2a}
\\
L^{0\,1}_{0\,2,-} &= L^{0\,2}_{0\,1,-}/3=L^{0\,3}_{0\,0,-}/6 \, ,\nonumber
%\label{eq3a}
\\
L^{2\,1}_{0\,0,+} &= 0\ ,\nonumber
%\label{eq4}
\\
L^{0\,1}_{2\,0,+} &= L^{1\,1}_{1\,0,+}=2 \, L^{1\,0}_{1\,1,+}-L^{2\,0}_{0\,1,+}\ ,\nonumber
%\label{eq5}
\\
L^{2\,0}_{0\,1,-} &= L^{1\,1}_{1\,0,-} = L^{2\,1}_{0\,0,-}/2 
= 
L^{0\,1}_{2\,0,-} + 2 \, L^{1\,0}_{1\,1,-} \, ,
\nonumber \\
%\label{eq6}
%\\
L^{1\,2}_{0\,0,+} &= 0\ ,\nonumber
%\label{eq4a}
\\
L^{1\,0}_{0\,2,+} &= L^{1\,1}_{0\,1,+}=2 \, L^{0\,1}_{1\,1,+}-L^{0\,2}_{1\,0,+}\ ,\nonumber
%\label{eq5a}
\\
L^{0\,2}_{1\,0,-} &= L^{1\,1}_{0\,1,-} = L^{1\,2}_{0\,0,-}/2 
= 
L^{1\,0}_{0\,2,-} + 2 \, L^{0\,1}_{1\,1,-} \, . 
\label{eq6a}
\end{align}
%----------------------------------------------------------

\section{The minimal-correlation coordinate}
\label{mcc}

Here we provide more explanations  concerning  the hidden redundancy  in the expressions for the  transport coefficients.

A ubiquitous  procedure to characterize the linear-response conductances of a three-terminal junction is to introduce a
 $3 \times 3$ conductance matrix $K_{rr'}$, in terms of which the energy currents are
%----------------------------------------------------------
\begin{align}
I_{Er}
=
\sum_{r'}
K_{r r'}
{\mathcal A}_{r'}
\, .\label{const}
\end{align}
%----------------------------------------------------------
The heat conductance matrix thus includes nine elements, but these are not independent. For example, when that matrix  is derived from Eq. (\ref{cgfsm}) upon using the identity $(d/dx)\ln{\rm det}{\mathbf M}={\rm Tr}\{{\mathbf M}^{-1}d{\mathbf M}/dx\}$ (where ${\mathbf M}$ is an arbitrary matrix), one finds
%----------------------------------------------------------
\begin{align}
K_{r r'}
&=
\left.
\frac{\partial^2 
{\mathcal F}_G(\{ \lambda_r \}, \{ {\mathcal A}_r \};B)
}
{
\partial {\mathcal A}_{r'}
\partial (i \lambda_r)
}
\right|_{
\lambda_r={\mathcal A}_{r}=0
}
\nonumber\\
&=
\frac{1}{4 \pi}
\int d \omega
\, 
\omega^2 \, 
\frac{
\delta_{rr'} - |S_{rr'}(\omega;B)|^2
}{1+\cosh(\beta \omega)}
\, \ .
\label{linheacon}
\end{align}
%----------------------------------------------------------
From  the unitarity of the scattering matrix  ${\mathbf S}$  it follows that  
%----------------------------------------------------------
\begin{align}
\sum_r
K_{rr'}
=
\sum_{r'}
K_{rr'}
=0
\, , 
\label{conlaw}
\end{align}
%----------------------------------------------------------
and therefore  the nine components of ${\mathbf K}$ are not independent. 
The transport coefficients  as  introduced in Eq.~(\ref{nltrc}) are free of redundancies and provide compact expressions.

There are many ways to remove the above-mentioned redundancies. 
For example, in deriving Eq. (\ref{choi}) we have assigned the same asymmetry parameter (denoted $x$) to the counting fields and to the affinities. 
We could have chosen a different asymmetry parameter for the counting fields and for the affinities, and rewrite Eq. (\ref{choi}) in a general form
%----------------------------------------------------------
\begin{align}
&{\mathcal F}^{}_G(\lambda,\lambda^{}_P,{\mathcal A},{\mathcal A}^{}_P;B,x,y)
\nonumber \\
&\equiv
{\mathcal F}^{}_G( (1+y) \lambda, y \lambda, \lambda^{}_P, 
(1+x) {\mathcal A}, x {\mathcal A}, {\mathcal A}^{}_P
;B)
\, . 
\label{choiy}
\end{align}
%----------------------------------------------------------
Similar to Eq.~(\ref{nltrc}), the transport coefficients in the present scheme are given by
%----------------------------------------------------------
\begin{align}
\label{nltrcy}
&L^{j \ell}_{k m}(B,x,y)
\\
&\equiv
\left.
\frac{\partial^{j+k+\ell +m} {\mathcal F}_G
(\lambda,\lambda_P,{\mathcal A},{\mathcal A}_P;B,x,y)
}
{
\partial \lambda^j
\partial {\mathcal A}^k
\partial \lambda_{P}^\ell
\partial {\mathcal A}_{P}^m
}
\right|_{\lambda=\lambda_{P}={\mathcal A}={\mathcal A}_{P}=0}
\, .\nonumber
\end{align}
%----------------------------------------------------------
These transport coefficients are related to the heat conductance matrix Eq. (\ref{linheacon}) via  
%----------------------------------------------------------
\begin{align}
L^{10}_{10}(B,x,y)
=&
K_{LL}
+
(K_{LL}+K_{RL}) \, y
+
(K_{LL}+K_{LR}) \, x
\nonumber \\
&+
(K_{LL}+K_{LR}+K_{RL}+K_{RR}) \, xy
\, ,
\nonumber \\
L^{10}_{01}(B,x,y)
=&
K_{LP}
+
(K_{LP}+K_{RP}) \, y
\, ,
\nonumber \\
L^{01}_{10}(B,x,y)
=&
K_{PL}
+
(K_{PL}+K_{PR}) \, x
\, ,
\nonumber \\
L^{01}_{01}(B,x,y)
=&
K_{PP}
\, . 
\label{trans}
\end{align}
%----------------------------------------------------------
In the approach of minimal-correlation coordinate, one chooses  $y$ as~\cite{Jordan} 
%----------------------------------------------------------
\begin{align}
y=
-
\frac{K_{LP}}{K_{LP}+K_{RP}}
\, .
\label{mincor}
\end{align}
%----------------------------------------------------------
Then, from the second of Eqs. ~(\ref{trans})
%----------------------------------------------------------
\begin{align}
L^{10}_{01}(B;-K_{LP}/(K_{LP}+K_{RP}))=0 \, .
\end{align}
%----------------------------------------------------------
This choice removes many of the vertex corrections discussed in Sec. \ref{subvc} (see Figs. \ref{diagnoise}, \ref{diagg2},  and \ref{diags1} there). 
Especially, for the linear transport case, vertex corrections vanish (Fig. \ref{diagnoise}).

\section{The third cumulant of the two-terminal junction}
\label{3d}

The third derivative of     ${\mathcal F}$, the  
 CGF of the effective  two-terminal junction,  is obtained by taking the complete derivative of Eq. (\ref{2ndd}) and exploiting  Eq.~(\ref{evc}).  
We find 
($\gamma,\gamma',\gamma''=c,q$) 
%----------------------------------------------------------
\begin{widetext}
\begin{align}
\frac{d^3 {\cal F}}{d a_{\gamma''}^{}\, d a_{\gamma'}^{} \, d a_{\gamma}^{}}
=&
\frac{\partial^3 {\mathcal F}^{}_G}{\partial a_{\gamma''}^{}\, \partial a_{\gamma'}^{}\, \partial a_{\gamma}^{}}
\nonumber \\
&
+
\sum_{\alpha,\alpha'=c,q}
\frac{\partial^3 {\mathcal F}^{}_G}{\partial a_{\gamma''}^{} \, \partial a_{\gamma'}^{}\, \partial v^{}_{\alpha}}
\,
U_{\alpha \alpha'}
\,
\frac{\partial^2 {\mathcal F}^{}_G}{\partial a_{\gamma}^{} \, \partial v^{}_{\alpha'}}
+
\frac{\partial^2 {\mathcal F}^{}_G}{\partial a_{\gamma'}^{} \, \partial v^{}_{\alpha}}
\,
\frac{\partial U_{\alpha \alpha'}}{\partial a_{\gamma''}^{}}
\,
\frac{\partial^2 {\mathcal F}^{}_G}{\partial a_{\gamma}^{} \, \partial v^{}_{\alpha'}}
+
\frac{\partial^2 {\mathcal F}^{}_G}{\partial a_{\gamma'}^{} \, \partial v^{}_{\alpha}}
\,
U^{}_{\alpha \alpha'}
\,
\frac{\partial^3 {\mathcal F}^{}_G}
{\partial a_{\gamma''}^{} \, \partial a_{\gamma}^{} \, \partial v^{}_{\alpha'}}
\nonumber 
\\
&
+
\sum_{\beta , \beta'=c,q}
\frac{\partial^2 {\mathcal F}^{}_G}{\partial a_{\gamma''}^{}\partial v^{}_{\beta'}}
U_{\beta' \beta}
\left \{
\frac{\partial^3 {\mathcal F}^{}_G}{\partial a_{\gamma'}^{} \, \partial a_{\gamma}^{} \, \partial v_{\beta}^{}}
\right.
\nonumber 
\\
&
+
\left.
\sum_{\alpha,\alpha'=c,q}
\left [
\frac{\partial^3 {\mathcal F}^{}_G}{\partial a_{\gamma'}^{} \, \partial v^{}_{\beta} \, \partial v^{}_{\alpha}}
\,
U^{}_{\alpha \alpha'}
\,
\frac{\partial^2 {\mathcal F}^{}_G}{\partial a_{\gamma}^{} \, \partial v^{}_{\alpha'}}
+
\frac{\partial^2 {\mathcal F}^{}_G}{\partial a_{\gamma'}^{} \, \partial v^{}_{\alpha}}
\,
\frac{\partial U^{}_{\alpha \alpha'} }{\partial v^{}_{\beta}}
\,
\frac{\partial^2 {\mathcal F}^{}_G}{\partial a_{\gamma}^{}\, \partial v^{}_{\alpha'}}
+
\frac{\partial^2 {\mathcal F}^{}_G}{\partial a_{\gamma'}^{} \, \partial v^{}_{\alpha}}
\,
U^{}_{\alpha \alpha'}
\,
\frac{\partial^3 {\mathcal F}^{}_G}{\partial a_{\gamma}^{} \, \partial v^{}_{\beta} \, \partial v^{}_{\alpha'}}
\right ]
\right \}
\nonumber\\
=&
\frac{\partial^3 {\mathcal F}^{}_G}{\partial a_{\gamma''}^{}\partial a_{\gamma'}^{} \partial a_{\gamma}^{}}
+
\sum_{ \{ i, j,k \} }
\sum_{\alpha,\alpha'=c,q}
\frac{\partial^2 {\mathcal F}^{}_G}{\partial a_{i}^{} \partial v^{}_{\alpha'}}
\,
U^{}_{\alpha' \alpha}
\,
\frac{\partial^3 {\mathcal F}^{}_G}{\partial a_{j}^{} \partial a_{k}^{} \partial v^{}_{\alpha}}
\nonumber \\
&
+
\sum_{ \{ i,j,k \} }
\sum_{\alpha_1,\alpha_2,\alpha_1',\alpha_2'=c,q}
\frac{\partial^2 {\mathcal F}^{}_G}
{\partial a_{i}^{} \partial v^{}_{\alpha_1'}}
\,
U_{\alpha_1' \alpha_{1}^{}}
\,
\frac{\partial^2 {\mathcal F}^{}_G}
{\partial a_{j }^{}\partial v^{}_{\alpha_2'}}
\,
U_{\alpha_2' \alpha_{2}^{}}
\,
\frac{\partial^2 {\mathcal F}^{}_G}{
\partial a^{}_k 
\partial v^{}_{\alpha_1} 
\partial v^{}_{\alpha_2}
}
\nonumber \\
&+
\sum_{\alpha_1,\alpha_2,\alpha_3,\alpha_1',\alpha_2',\alpha_3'=c,q}
\frac{\partial^2 {\mathcal F}^{}_G}{\partial a_{\gamma}^{} \partial v^{}_{\alpha_1'}}
\,
U^{}_{\alpha_1' \alpha_1}
\,
\frac{\partial^2 {\mathcal F}^{}_G}{\partial a_{\gamma'}^{} \partial v^{}_{\alpha_2'}}
\,
U_{\alpha_{2}' \alpha_{2}^{}}
\,
\frac{\partial^2 {\mathcal F}^{}_G}{\partial a_{\gamma''}^{} \partial v^{}_{\alpha_3'}}
\,
U_{\alpha_{3}' \alpha_{3}^{}}
\,
\frac{\partial^3 {\mathcal F}^{}_G}{
\partial v^{}_{\alpha_1} 
\partial v^{}_{\alpha_2}
\partial v^{}_{\alpha_3} 
}
\, ,
\label{3rdd}
\end{align}
%----------------------------------------------------------
where we have used Eqs.  (\ref{dudv}). 
Here $\sum_{ \{ i,j,k \} }$ means the summation over the three cyclic combinations of $\{ \gamma, \gamma' ,\gamma'' \}$ .

\section{The symmetric and anti-symmetric components of the noise with respect to a magnetic field}
\label{mag}

As is explained in the main text, the full expression for the linear-response noise (see Fig.~\ref{diags1}) turns out to be rather cumbersome. However, it is possible to obtained certain relations among its symmetric and anti-symmetric components, with respect to the magnetic field. 

The symmetric component is 
%----------------------------------------------------------
\begin{align}
L^{1}_{2,+}
=&
L^{2}_{1,+}
=
L^{2 \, 0}_{1 \, 0,+}
+
2
\, 
\frac{
L^{1 \, 0}_{0 \, 1,+} 
(L^{2 \, 0}_{0 \, 1,+} - 4 L^{1 \, 0}_{1 \, 1,+})
-
L^{1 \, 0}_{0 \, 1,-} 
(L^{0 \, 1}_{2 \, 0,-}-2 L^{1 \, 0}_{1 \, 1,-})
}
{L^{0 \, 2}_{0 \, 0,+}}
\nonumber \\
&+
4 
\, 
\frac{ 
(L^{1 \, 0}_{0 \, 1,+})^2 
(L^{0 \, 2}_{1 \, 0,+}+2 L^{1 \, 1}_{0 \, 1,+})
-
(L^{1 \, 0}_{0 \, 1,-})^2 
L^{0 \, 2}_{1 \, 0,+}
-
2 
L^{1 \, 0}_{0 \, 1,+}
L^{1 \, 0}_{0 \, 1,-}
L^{1 \, 0}_{0 \, 2,-}
}
{ (L^{0 \, 2}_{0 \, 0,+})^2 }
\nonumber \\
&
-
8 
\, 
\frac{
[(L^{1 \, 0}_{0 \, 1,-})^2 - (L^{1 \, 0}_{0 \, 1,+})^2]
(
L^{1 \, 0}_{0 \, 1,-} L^{0 \, 1}_{0 \, 2,-}
-
L^{1 \, 0}_{0 \, 1,+} L^{0 \, 2}_{0 \, 1,+}
)
}
{(L^{0 \, 2}_{0 \, 0,+})^3}
\, ,
\label{magsym}
\end{align}
%----------------------------------------------------------
and the asymmetric component is 
%----------------------------------------------------------
\begin{align}
L^{1}_{2,-}
=&
L^2_{1,-}/3
=
L^{1 \, 0}_{2 \, 0,-}
+
2 \, 
\frac{
L^{1 \, 0}_{0 \, 1,-}  L^{2 \, 0}_{0 \, 1,+}
- 
L^{1 \, 0}_{0 \, 1,+} 
(L^{0 \, 1}_{2 \, 0,-} + 2 L^{1 \, 0}_{1 \, 1,-})
}
{L^{0 \, 2}_{0 \, 0,+}}
\nonumber \\
&+
4 \, 
\frac{ 
2 [
(L^{1 \, 0}_{0 \, 1,+})^2 
-
(L^{1 \, 0}_{0 \, 1,-})^2 
]
\, 
L^{0 \, 1}_{1 \, 1,-}
+
[
(L^{1 \, 0}_{0 \, 1,+})^2 
+
(L^{1 \, 0}_{0 \, 1,-})^2 
]
\, 
L^{1 \, 0}_{0 \, 2,-}
-
2 
L^{1 \, 0}_{0 \, 1,-}
L^{1 \, 0}_{0 \, 1,+}
L^{1 \, 1}_{0 \, 1,+}
}
{ (L^{0 \, 2}_{0 \, 0,+})^2 }
\nonumber \\
&
-
8 \, 
\frac{
[(L^{1 \, 0}_{0 \, 1,-})^2 - (L^{1 \, 0}_{0 \, 1,+})^2]
(
L^{1 \, 0}_{0 \, 1,-} L^{0 \, 2}_{0 \, 1,+}
-
L^{1 \, 0}_{0 \, 1,+} L^{0 \, 1}_{0 \, 2,-}
)
}
{(L^{0 \, 2}_{0 \, 0,+})^3}
\, . 
\label{magasym}
\end{align}
%----------------------------------------------------------

\end{widetext}

%\end{appendix}


\begin{thebibliography}{99}
\bibitem {com}
Note, though, that the problem considered here is somewhat simpler, since the dynamics of the  classical  variable, i.e., the energy of the probe, is not governed by a Hamiltonian.
\end{thebibliography}
\end{document}